\documentclass[iop]{emulateapj}

\newcommand{\kms}{\textrm{km~s$^{-1}$}}

\newcommand{\msolar}{M$_{\odot}$}
\newcommand{\ml}{M$_{\odot}$~yr$^{-1}$}

\usepackage{subfigure}
\usepackage{multirow}
\usepackage[dvips]{color}
\definecolor{Mygrey}{gray}{0.6}

\begin{document}

\title{Uncovering the Putative B-Star Binary Companion of the SN 1993J Progenitor}
\shorttitle{SN 1993J Binary Companion}
\author{Ori D. Fox\altaffilmark{1,2}, K. Azalee Bostroem\altaffilmark{3}, Schuyler D. Van Dyk\altaffilmark{4}, Alexei V. Filippenko\altaffilmark{1}, Claes Fransson\altaffilmark{5}, Thomas Matheson\altaffilmark{6}, S. Bradley Cenko\altaffilmark{1,7}, Poonam Chandra\altaffilmark{8}, Vikram Dwarkadas\altaffilmark{9}, Weidong Li\altaffilmark{1,10}, Alex H. Parker\altaffilmark{1}, and Nathan Smith\altaffilmark{11}}
\altaffiltext{1}{Department of Astronomy, University of California, Berkeley, CA 94720-3411, USA.}
\altaffiltext{2}{ofox@berkeley.edu.}
\altaffiltext{3}{Space Telescope Science Institute, 3700 San Martin Drive, Baltimore, MD 21218, USA.}
\altaffiltext{4}{Caltech, Mailcode 314-6, Pasadena, CA 91125, USA.}
\altaffiltext{5}{Department of Astronomy, Oskar Klein Centre, Stockholm University, AlbaNova,
SE--106~91 Stockholm, Sweden.} 
\altaffiltext{6}{National Optical Astronomy Observatory, 950 North Cherry Avenue, Tucson, AZ 85719-4933, USA.}
\altaffiltext{7}{Astrophysics Science Division, NASA Goddard Space Flight Center, Mail Code 661, Greenbelt, MD 20771, USA.}
\altaffiltext{8}{National Centre for Radio Astrophysics, Tata Institute of Fundamental Research, Pune University Campus, Ganeshkhind, Pune-411007, India.}
\altaffiltext{9}{Department of Astronomy and Astrophysics, University of Chicago, 5640 S Ellis Ave, Chicago, IL 60637. }
\altaffiltext{10}{Deceased 2011 December 12.}
\altaffiltext{11}{Steward Observatory, 933 N. Cherry Ave., Tucson, AZ 85721, USA.}

\begin{abstract}

The Type IIb supernova (SN) 1993J is one of only a few stripped-envelope supernovae with a progenitor star identified in pre-explosion images.  
SN~IIb models typically invoke H envelope stripping by mass transfer in a binary system.  For the case of SN 1993J, the models suggest that the 
companion grew to 22 \msolar\ and became a source of ultraviolet (UV) excess.  Located in M81, at a distance of only 3.6 Mpc, SN 1993J 
offers one of the best opportunities to detect the putative companion and test the progenitor model.  Previously published near-UV spectra in 2004 showed evidence for absorption lines consistent with a hot (B2~Ia) star, but the field was crowded and dominated by flux from the SN.  Here we present {\it Hubble Space Telescope (HST)} Cosmic Origins Spectrograph (COS) and Wide-Field Camera 3 (WFC3) observations of SN 1993J  from 2012, at which point the flux from the SN had faded sufficiently to potentially measure the UV continuum properties from the putative companion.  The resulting UV spectrum is consistent with contributions from both a hot B star and the SN, although we cannot rule out line-of-sight coincidences.

\end{abstract}


\section{Introduction}
\label{sec:intro}

The Type IIb supernova (SN~IIb; see \citealt{filippenko97} for a review) subclass exhibits spectroscopic evidence for hydrogen emission at early times (the defining property of Type II supernovae, SNe~II), He~I absorption at later times (as in SNe~Ib), and renewed broad H$\alpha$ emission in the nebular phase \citep[e.g.,][]{filippenko88,matheson00,taubenberger11,shivvers13}.  While SNe~IIb require most of the progenitor's original H envelope to be lost prior to the SN explosion, a low-mass (0.01--0.5 \msolar) H envelope must be retained.  Single massive stars (e.g., $>30$ \msolar) could potentially lose their  envelopes in a stellar wind, but models indicate that only a limited range of initial masses would result in a very low-mass envelope  at the time of the explosion \citep[][and references therein]{podsiadlowski93, woosley94}.  More likely progenitor scenarios involve  lower-mass stars (e.g., evolved red supergiants) that lose their envelopes during mass transfer to a binary companion \citep{podsiadlowski93,nomoto93,woosley94}.  The binary progenitor scenario is also consistent with SN rates, which strongly suggest that a substantial fraction of stripped-envelope SNe, including SNe~IIb, must originate with relatively lower-mass stars ($<$25 \msolar~initially) stripped in binary systems \citep{smith11}.

Along with SNe 2013df \citep{vandyk13b}, 2011dh \citep{maund11,vandyk11,ergon13,vandyk13}, and 2008ax \citep{li08,crockett08}, SN 1993J is one of just four SNe~IIb with progenitors directly identified in pre-explosion images \citep{aldering94,cohen95}.  The progenitor star, a K supergiant, was consistent with SN~IIb models that invoke supergiant mass loss onto a binary companion.  By 2006, the SN had faded sufficiently for \citet{maund09} to confirm that the K-supergiant signature in the spectral energy distribution (SED) had disappeared. Even before this disappearance, however, excess flux in the near-ultraviolet (NUV) and $B$ bands suggested the possible presence of the hot companion, thereby supporting the binary-star model \citep[e.g.,][]{vandyk02,maund04}.  

%% Table of HST Photometric Observations
\begin{deluxetable*}{ l c c c c c}
\tablewidth{0pt}
\tabletypesize{\normalsize}
\tablecaption{$HST$~GO-12531 COS/WFC3 Photometric Observations \label{tab1}}
\tablecolumns{6}
\tablehead{
\multirow{2}{*}{UT Date} & \colhead{Epoch} &\multirow{2}{*}{Instrument} & \multirow{2}{*}{Filter} & \colhead{Central Wavelength} & \colhead{Exposure}\\
                                            &\colhead{(days)}   &                                                &                                        & \colhead{(\AA)}                              &\colhead{(s)}              
}
\startdata
20120331 & 6944 & COS & MIRRORA/G140L & 1105 & 7466 \\
20120406 & 6950 & COS & MIRRORA/G230L & 2950 & 33752 \\
20120411 & 6955 & COS & MIRRORA/G230L & 2635 & 13651 \\
20120411 & 6955 & COS & MIRRORA/G230L & 3360 &  20101\\
20111226 & 6848 & WFC3/UVIS\tablenotemark{1} & F218W & 2228.5 & 3000 \\
20120218 & 6902 & WFC3/UVIS & F275W & 2710.2 & 3000 \\
20120219 & 6903 & WFC3/UVIS & F336W & 3354.8 & 3000 \\
20111224 & 6846 & WFC3/UVIS & F438W & 4326.5 & 856\\
20111224 & 6846 & WFC3/UVIS & F555W & 5308.1 & 856\\
20111224 & 6846 & WFC3/UVIS & F625W & 6241.3 & 856\\
20120219 & 6903 & WFC3/UVIS & F814W & 8029.5 & 836\\
20120219 & 6903 & WFC3/UVIS & F850LP & 9168.80  & 836\\
20120219 & 6903 & WFC3/UVIS & F105W & 10552.00 & 328\\
20120219 & 6903 & WFC3/UVIS & F125W & 12486.00 & 328\\
20120219 & 6903 & WFC3/UVIS & F160W & 15369.00 & 328
\enddata
\tablenotetext{1}{For the UVIS images, the UVIS2 aperture was used, so that the SN field was near the edge of chip 2.  The UVIS exposures were dithered. No post-flash was used for any of this imaging.  The IR observations in each band were conducted using the SPARS25 sequence with the number of samples equal to 14.}
\end{deluxetable*}

\begin{figure}[h]
\vspace{0.2in}
%\begin{landscape}
\begin{center}
\epsscale{0.55}
\subfigure{\label{f1a} \plotone{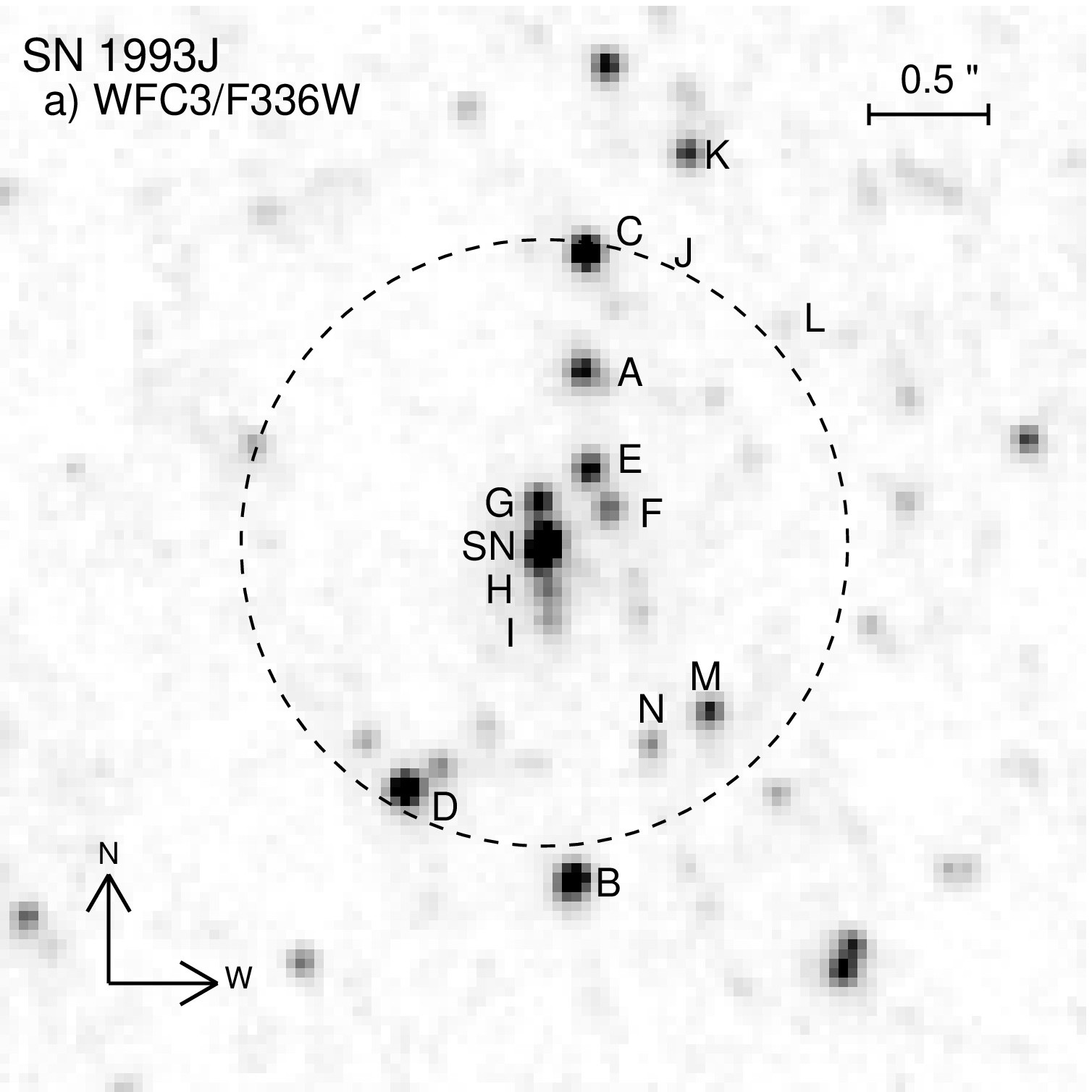}}
\subfigure{\label{f1b} \plotone{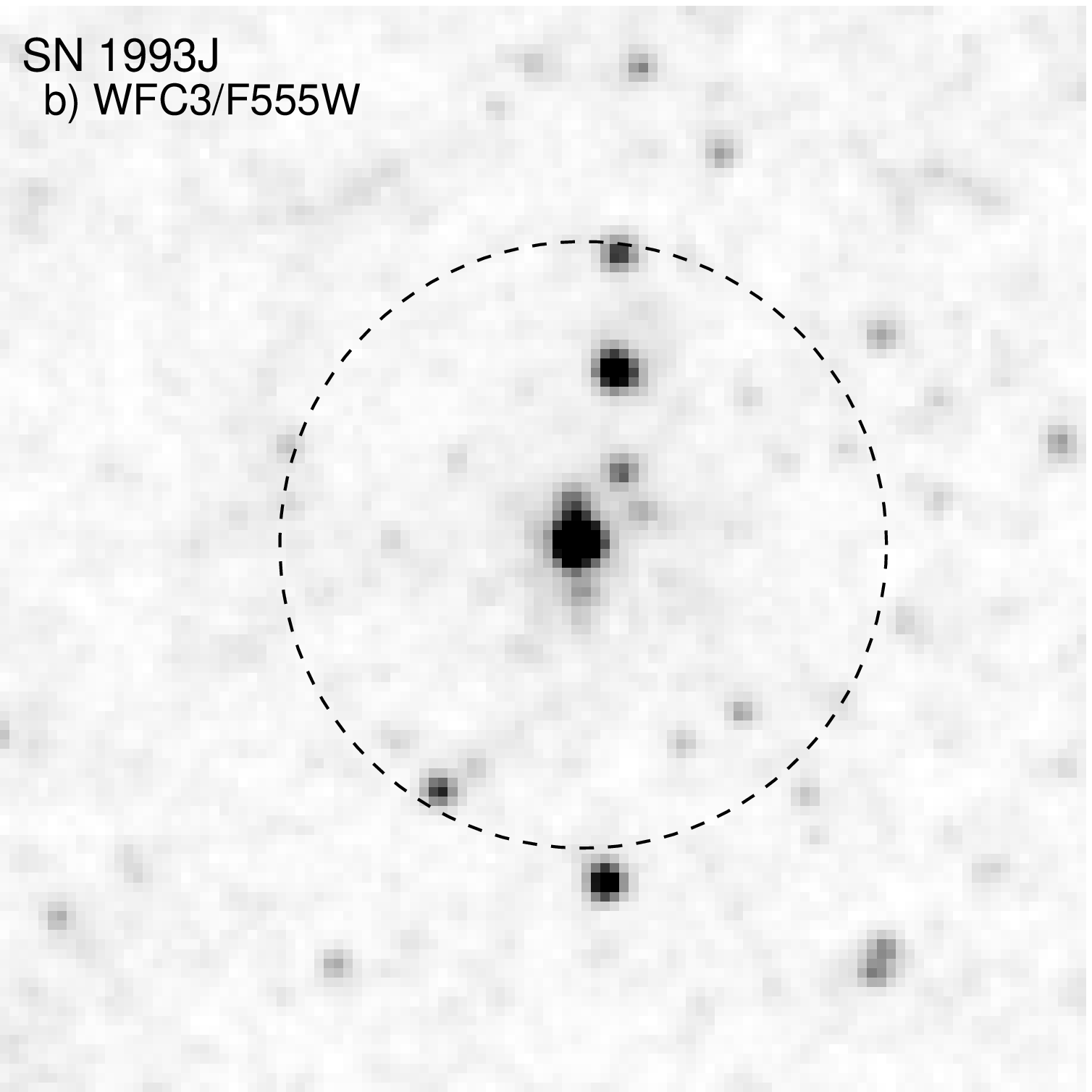}}\\
\subfigure{\label{f1c} \plotone{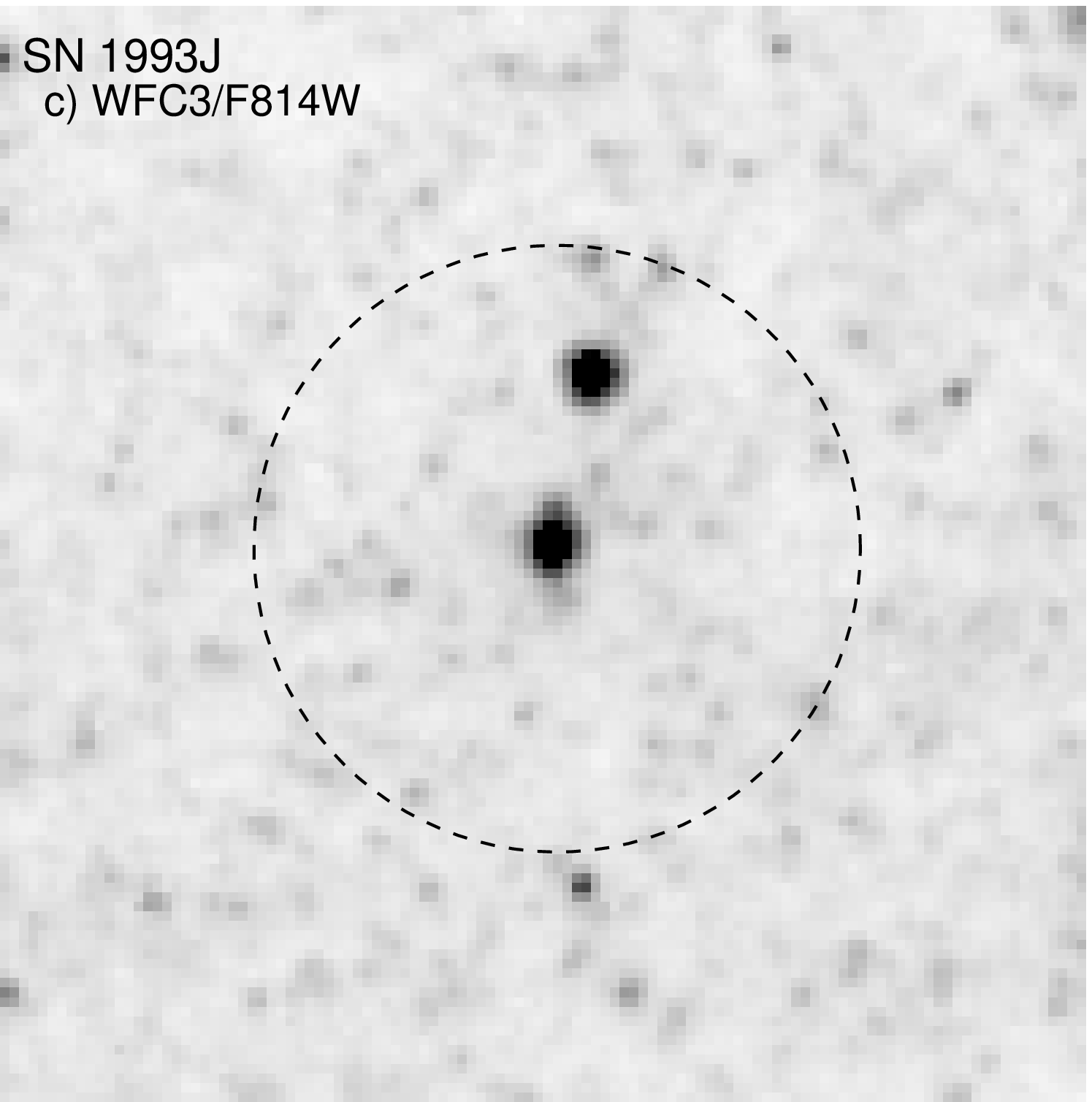}}
\subfigure{\label{f1d} \plotone{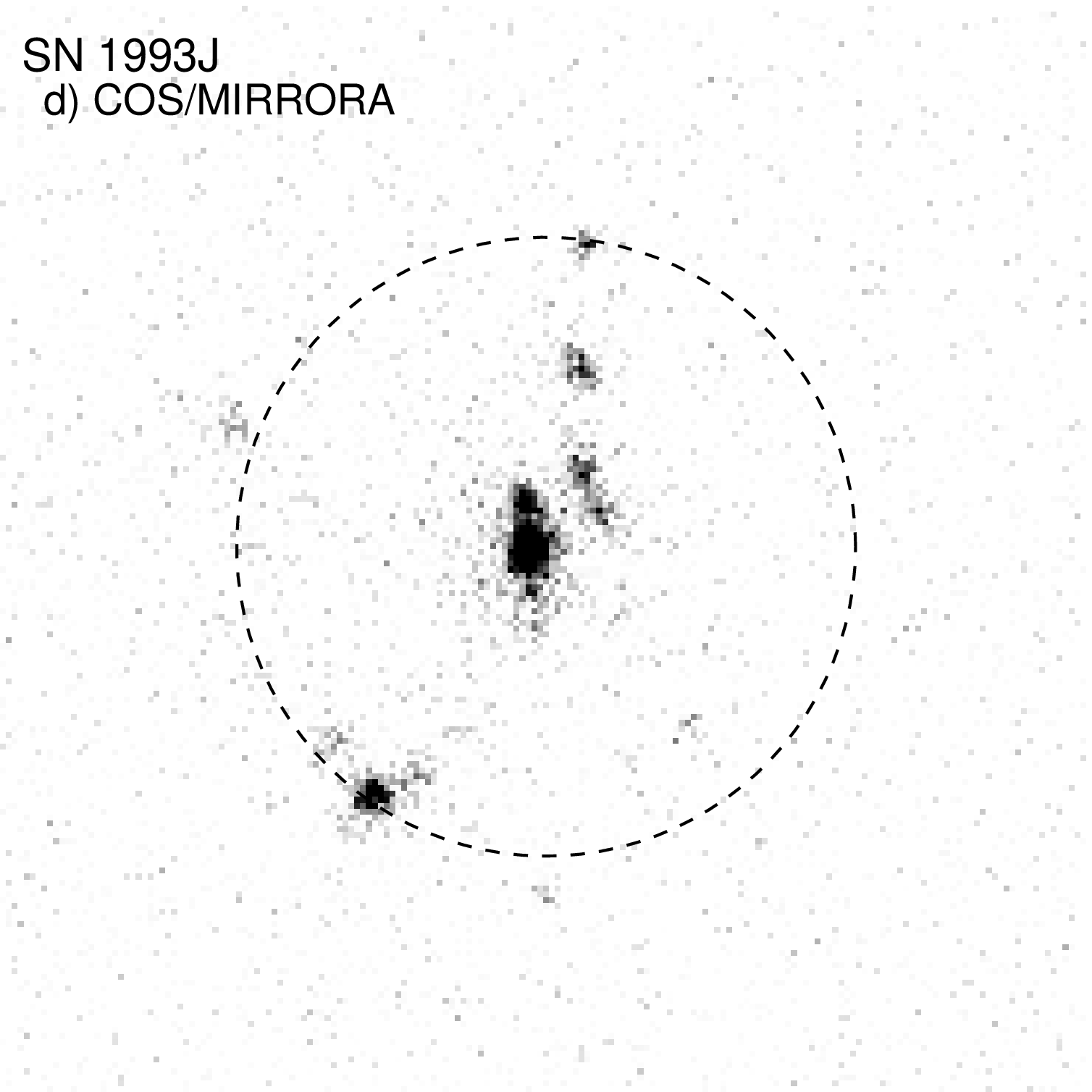}}
\caption{Field of SN 1993J imaged in (a) WFC3/F336W, (b) WFC3/F555W, (c) WFC3/F814W, and (d) COS NUV, all to approximately the same scale and orientation.  The dashed black circles correspond to the COS 2.5\arcsec-diameter aperture, which can also be considered poor seeing from the ground. Star labels designate the SN and other stars previously identified in {\it HST\/} images by \citet{vandyk02} and \citet{maund04}, together with additional fainter stars.
}
\label{fig1}
\end{center}
%\end{landscape}
\vspace{0.2in}
\end{figure}

But the specific source of the NUV excess was not immediately obvious.  SN 1993J has a number of stars within 1\arcsec\ of it (Fig. \ref{fig1}).  Using $HST$ images, \citet{vandyk02} initially showed that some, {\it but not all}, of the excess NUV could be explained by previously unresolved stars.  \citet{maund04} later obtained an optical/NUV spectrum of SN 1993J and the putative companion with the Low Resolution Imaging Spectrometer (LRIS; Oke et al. 1995) on the Keck-I telescope and detected spectral absorption features consistent with a massive B-type supergiant.  Even with a 0.7\arcsec-wide slit, however, these spectra would have been contaminated by Stars E, F, G, H, and I (Fig. \ref{fig1}).  \citet{maund04} calculated an upper limit on the contribution from the surrounding stars by simulating the spectral profile and concluded that the spectral features are dominated by a source at the position of the SN.  Line-of-sight coincidences were not ruled out.

While these results further support the binary scenario for SN 1993J, a {\it direct} detection of continuum from a hot B-star companion remains unconfirmed to date.  \citet{maund09} predicted that with the observed rate of SED decline, a B star at the location of SN 1993J could be 
detected in the $U$ and $B$ bands by 2012.  Here we present {\it HST\/} UV, optical, and infrared (IR) photometry and spectroscopy of SN 1993J obtained in November 2011 through April 2012.  Section \ref{sec:obs} summarizes the optical/IR photometry, \S \ref{sec:keck} presents our Keck spectrum, and \S \ref{sec:cos} describes the UV spectroscopy and spectral extraction procedures.  We analyze the data in \S \ref{sec:disc} with B-star models.  Our results and conclusions are discussed in \S \ref{sec:con}.

%FOVs

\section{WFC3 Observations}
\label{sec:obs}

SN 1993J was observed with the {\it HST} Wide-Field Camera 3 (WFC3) UVIS and IR channels in November 2011--February 2012 as part of program GO-12531 (PI A. Filippenko), as summarized in Table \ref{tab1}.  Figures \ref{fig1}(a) through \ref{fig1}(c) display the WFC3/F336W, F555W, and F814W images. Star labels those of \citet{vandyk02} and \citet{maund04}.

\begin{figure}
\vspace{0.3in}
\begin{center}
\hspace{-0.3in}\includegraphics[width=0.5\textwidth]{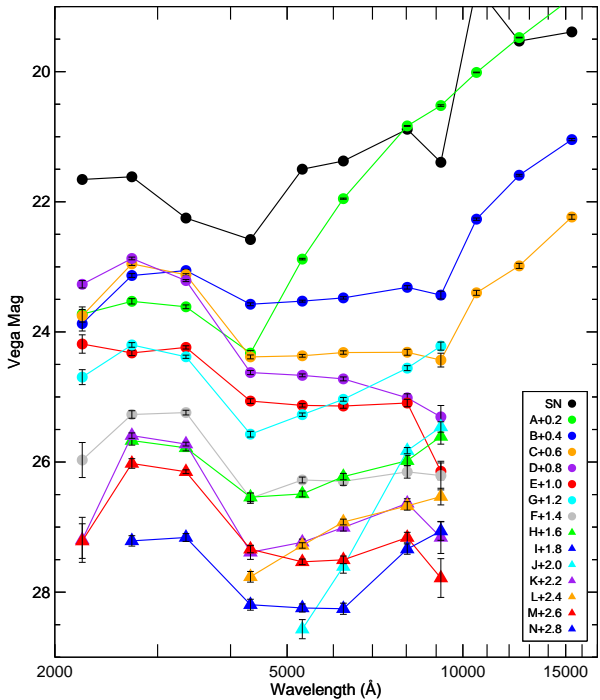}
\caption{WFC3 photometry of SN~1993J and the surrounding Stars A through N identified in Figure \ref{fig1}.
}
\vspace{-0.3in}
\label{fig2}
\end{center}
\end{figure}

Photometry was extracted from the individual WFC3 ``flt'' images in all bands using Dolphot v2.0 \citep{dolphin00}.  The input parameters are those recommended by the Dolphot WFC3 Users' Manual.  Aperture corrections were applied.  The F438W, F555W, and F625W data, all obtained on 2011 December 24 in a set of orbits and therefore with the same orientation, were processed together.  All other bands were obtained at other orientations and hence were processed in separate Dolphot runs.  The data acquired on 2012 February 19 in the F814W and F850LP bands exhibit residuals created by a bright (red) source or sources likely observed immediately prior.  The backgrounds around the SN environment in each of these bands are slightly elevated relative to the overall backgrounds in these images.  Since Dolphot was set to measure the sky background locally, we do not expect these elevated backgrounds to significantly affect the resulting photometry.  The F336W data obtained in the same sequence on that date appear not to show residuals.  The resulting magnitudes in the WFC3 flight system (Vegamag) are listed in Tables~\ref{tab2} and \ref{tab3} and plotted in Figure \ref{fig2}.

%% Table of WFC3 Photometry UVIS
\begin{deluxetable*}{ l c c c c c c c c}
\tablewidth{0pt}
\tabletypesize{\tiny}
%\rotate
\tablecaption{WFC3 UVIS Photometry \label{tab2}}
\tablecolumns{9}
\tablehead{
\colhead{Target} & \colhead{F218W} & \colhead{F275W} & \colhead{F336W} & \colhead{F438W} & \colhead{F555W} & \colhead{F625W} & \colhead{F814W} & \colhead{F850LP} \\
\colhead{} & \multicolumn{8}{c}{Vegamag (1$\sigma$ uncertainty)}
}
\startdata
93J    & 21.657 (0.037) & 21.616  (0.017) & 22.250 (0.016) & 22.580 (0.017) & 21.499 (0.006)) & 21.373 (0.007)) & 20.885 (0.009)) & 21.393 (0.022) \\ 
A        & 23.527 (0.111) & 23.330 (0.041) & 23.414 (0.029) & 24.126 (0.039) & 22.681 (0.012) & 21.752 (0.009) & 20.634 (0.007) & 20.321 (0.012)  \\ 
B        & 23.472 (0.114) & 22.733 (0.029) & 22.657 (0.018) & 23.175 (0.022) & 23.127 (0.015) & 23.077 (0.018) & 22.915 (0.026) & 23.043 (0.060)  \\ 
C        & 23.152 (0.097) & 22.354 (0.024) & 22.522 (0.017) & 23.784 (0.030) & 23.767 (0.021) & 23.717 (0.026) & 23.713 (0.045) & 23.833 (0.106)  \\ 
D        & 22.468 (0.057) & 22.071 (0.021) & 22.411 (0.016) & 23.825 (0.031) & 23.868 (0.022) & 23.922 (0.030) & 24.213 (0.062) & 24.506 (0.175)  \\ 
E        & 23.186 (0.141) & 23.323 (0.041) & 23.238 (0.025) & 24.063 (0.036) & 24.129 (0.026) & 24.139 (0.034) & 24.093 (0.057) & 24.143 (0.129)  \\ 
F        & 24.569 (0.269) & 23.869 (0.058) & 23.840 (0.036) & 25.157 (0.077) & 24.876 (0.045) & 24.897 (0.061) & 24.751 (0.096) & 24.808 (0.220)  \\ 
G       & 23.495 (0.115) & 23.000 (0.036) & 23.182 (0.025) & 24.372 (0.044) & 24.072 (0.026) & 23.836 (0.029) & 23.357 (0.038) & 23.018 (0.062) \\ 
H        & $>$25.0 & 24.070 (0.070) & 24.182 (0.047) & 24.940 (0.065) & 24.890 (0.045) & 24.624 (0.049) & 24.379 (0.078) & 24.007 (0.118) \\ 
I         &  $>$25.0 & 24.396 (0.083) & 24.532 (0.057) & 25.954 (0.132) & 25.661 (0.074) & 25.594 (0.098 ) &  $>$26.2 &  $>$26.4  \\ 
J        &  $>$25.0 &   $>$26.4 & $>$27.0 &  $>$27.2 & 26.569 (0.147) & 25.606 (0.102) & 23.825 (0.048) & 23.461 (0.080) \\ 
K       & 24.996 (0.346) & 23.394 (0.042) & 23.524 (0.029) & 25.187 (0.109) & 25.033 (0.048) & 24.805 (0.055) & 24.435 (0.071) & 24.959 (0.246)\\ 
L        &  $>$25.0  &  $>$26.4 &  $>$27.0 & 25.364 (0.082) & 24.882 (0.042) & 24.522 (0.044) & 24.274 (0.065) & 24.131 (0.128) \\ 
M       & 24.615 (0.266) & 23.422 (0.074) & 23.549 (0.029) & 24.740 (0.055) & 24.934 (0.043) & 24.903 (0.058) & 24.561 (0.081) & 25.183 (0.298) \\ 
N       &   $>$25.0 & 24.412 (0.080) & 24.360 (0.060) & 25.393 (0.084) & 25.443 (0.062) & 25.455 (0.086) & 24.538 (0.079) & 24.261 (0.142)
\enddata
\end{deluxetable*}

%% Table of WFC3 Photometry IR
\begin{deluxetable*}{ l c c c}
\vspace{-0.2in}
\tablewidth{0pt}
\tabletypesize{\normalsize}
\tablecaption{WFC3 IR Photometry\tablenotemark{1} \label{tab3}}
\tablecolumns{4}
\tablehead{
\colhead{Target} & \colhead{F105W} & \colhead{F125W} & \colhead{F160W}\\
\colhead{}   & \multicolumn{3}{c}{Vegamag (1$\sigma$ uncertainty)}
}
\startdata
SN 1993J/Companion & 18.691 (0.005) & 19.530 (0.011) & 19.389 (0.015)\\
A        & 19.812 (0.005) & 19.276 (0.005) & 18.666 (0.005)\\
B        & 21.867 (0.018) &  21.192 (0.014) &  20.644 (0.017)\\
C        & 22.799 (0.038) & 22.385 (0.039) & 21.634 (0.035)
\enddata
\tablenotetext{1}{Stars D--N were not detected by Dolphot and have the following 3$\sigma$~upper limits: F105W $>$ 24.8, F125W  $>$ 24.2, and F160W  $>$ 23.3 mag.}
\end{deluxetable*}

%Figure showing SED

%%Light Echoes
\begin{figure}[h]
%\begin{landscape}
\begin{center}
%\epsscale{0.45}
\subfigure{\label{f3a} \plotone{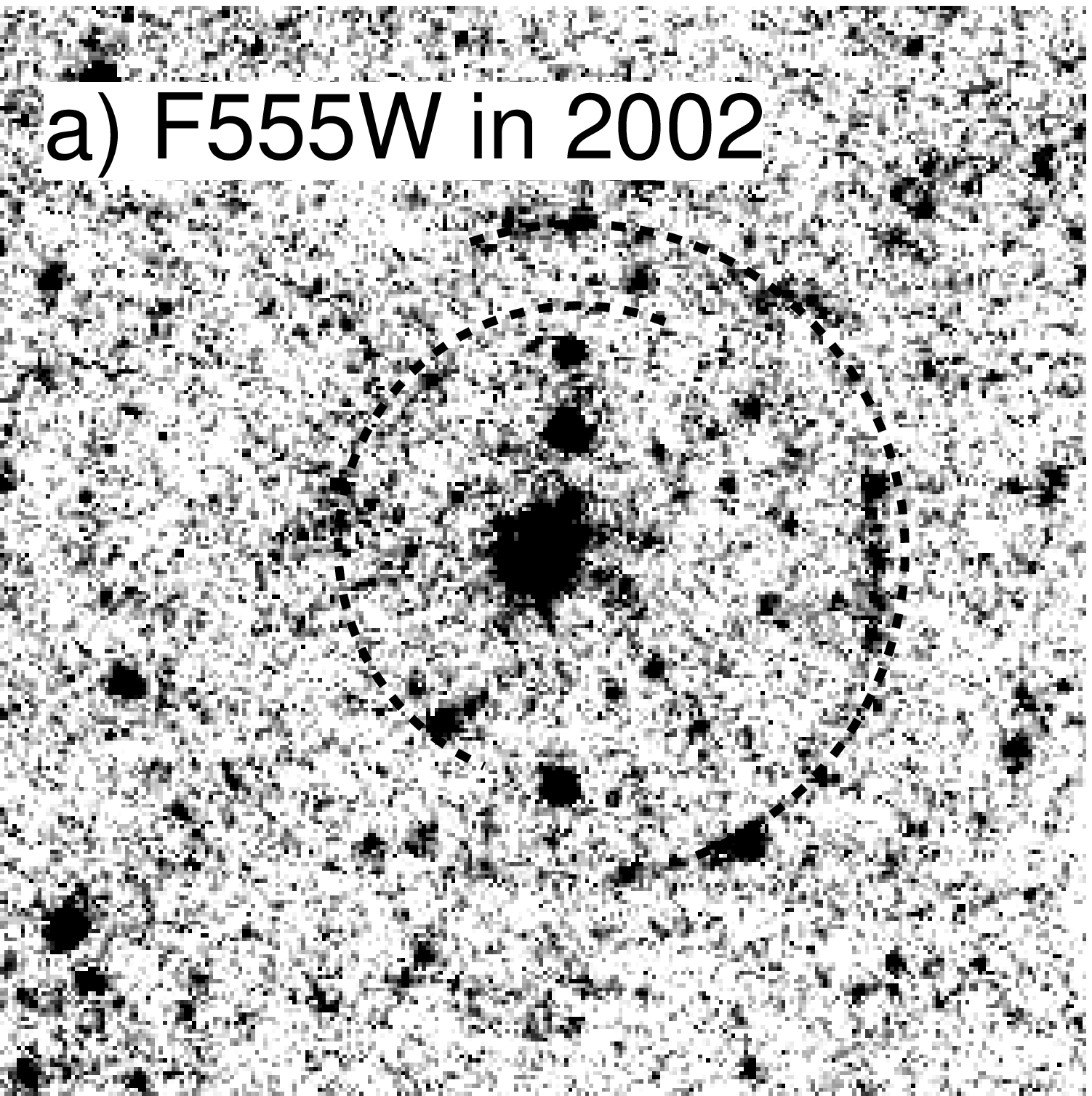}}\\
\subfigure{\label{f3b} \plotone{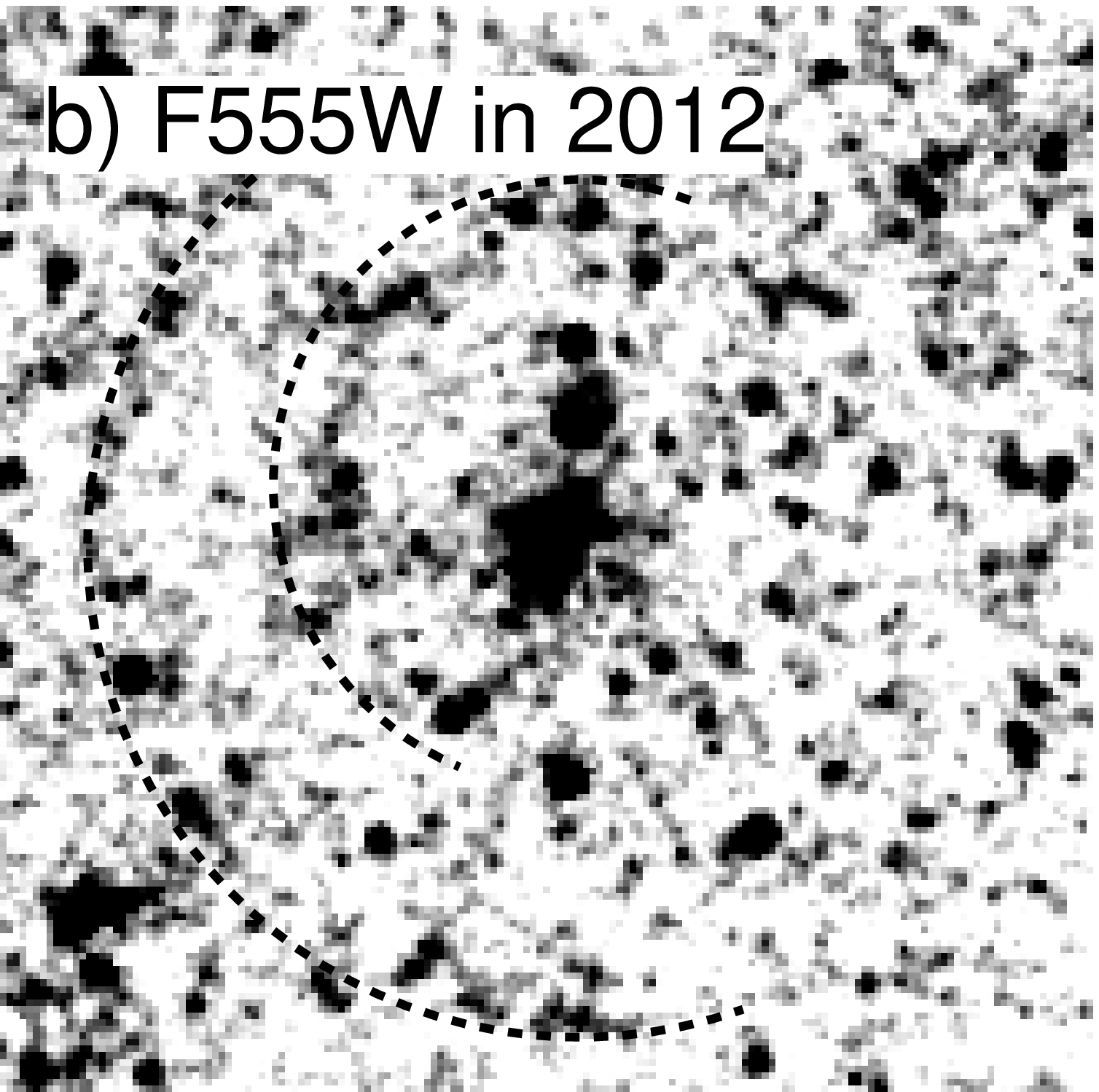}}\\
\caption{F555W images highlighting light echoes observed around SN 1993J in (a) 2002 and (b) 2012.  Both panels are showed in the same orientation as Figure \ref{fig1}, with North as up and West to the right.
}
\label{fig3}
\end{center}
%\end{landscape}
\end{figure}

We also detect the presence of expanding light echoes in the field of SN 1993J.  Figure \ref{fig3} shows F555W filtered images of SN 1993J taken in both 2002 and 2012.  The 2002 data were obtained with the $HST$~Advanced Camera for Surveys' (ACS) High Resolution Camera (HRC) and downloaded from the Hubble Legacy Archive (HLA)\footnote{http://hla.stsci.edu .}.  The comparison of these epochs highlights two expanding rings.  The expansion of these rings can be described similarly to echoes observed in other SNe, remnants, and flares \citep[e.g.,][and references within]{rest11,rest12}.  Owing to a time delay associated with distance, the observer witnesses the scattered SN light from different locations on large clouds of dust.  Here we highlight the detection of these echoes, but a complete analysis entails many subtleties and is beyond the scope of this paper.

\section{Keck Observations}
\label{sec:keck}

On 2013 February 17, we obtained a spectrum of SN 1993J with the DEep Imaging Multi-Object Spectrograph \citep[DEIMOS;][]{faber03}~mounted on the 10-m Keck~II telescope.  It was obtained using the 1200/7500 grating, along with a 1\arcsec-wide slit, resulting in a wavelength coverage of  4750--7400~\AA\ and a typical resolution of ~3~\AA.  The slit was aligned along the parallactic angle to minimize differential light losses \citep{filippenko82}.  

The data were reduced using standard techniques \citep[e.g.,][]{foley03,silverman12}. Routine CCD processing and spectrum extraction were completed with {\tt IRAF}, and the data were extracted with the optimal algorithm of \citet{horne86}. We obtained the wavelength scale from low-order polynomial fits to calibration-lamp spectra. Small wavelength shifts were then applied to the data after cross-correlating a template sky to an extracted night-sky spectrum. Using our own IDL routines, we fit a spectrophotometric standard-star spectrum to the data in order to flux calibrate the SN and to remove telluric absorption lines \citep{wade88,matheson00}.  

\section{COS Observations}
\label{sec:cos}

SN 1993J was observed with the {\it HST}~Cosmic Origins Spectrograph (COS) in March--April 2012 as part of program GO-12531 (PI A. Filippenko), as summarized in Table \ref{tab4}.  For all {\it HST} COS observations, SN 1993J was centered in the aperture using a COS NUV target acquisition (TA) image with MIRROR A and the Primary Science Aperture (PSA). Analysis of the TA images shows that this strategy was successful and that SN 1993J is centered in the aperture during all subsequent observations.  

%% Table of HST Spectroscopic Observations
\begin{deluxetable}{ l c c c c}
\tablewidth{0pt}
\tabletypesize{\normalsize}
\tablecaption{$HST$~GO-12531 COS Spectroscopic Observations \label{tab4}}
\tablecolumns{5}
\tablehead{
\multirow{2}{*}{UT Date} & \colhead{Epoch} & \multirow{2}{*}{Grating} & \colhead{Central Wavelength} & \colhead{Exposure}\\
			                &\colhead{(days)}   &                                           & \colhead{(\AA)}                               &\colhead{(s)}              
}
\startdata
20120331 & 6944 & G140L & 1105 & 7466 \\
20120406 & 6950 & G230L & 2950 & 33752 \\
20120411 & 6955 & G230L & 2635 & 13651 \\
20120411 & 6955 & G230L & 3360 & 20101
\enddata
\end{deluxetable}

\begin{figure*}
\includegraphics[width = 1.1\textwidth]{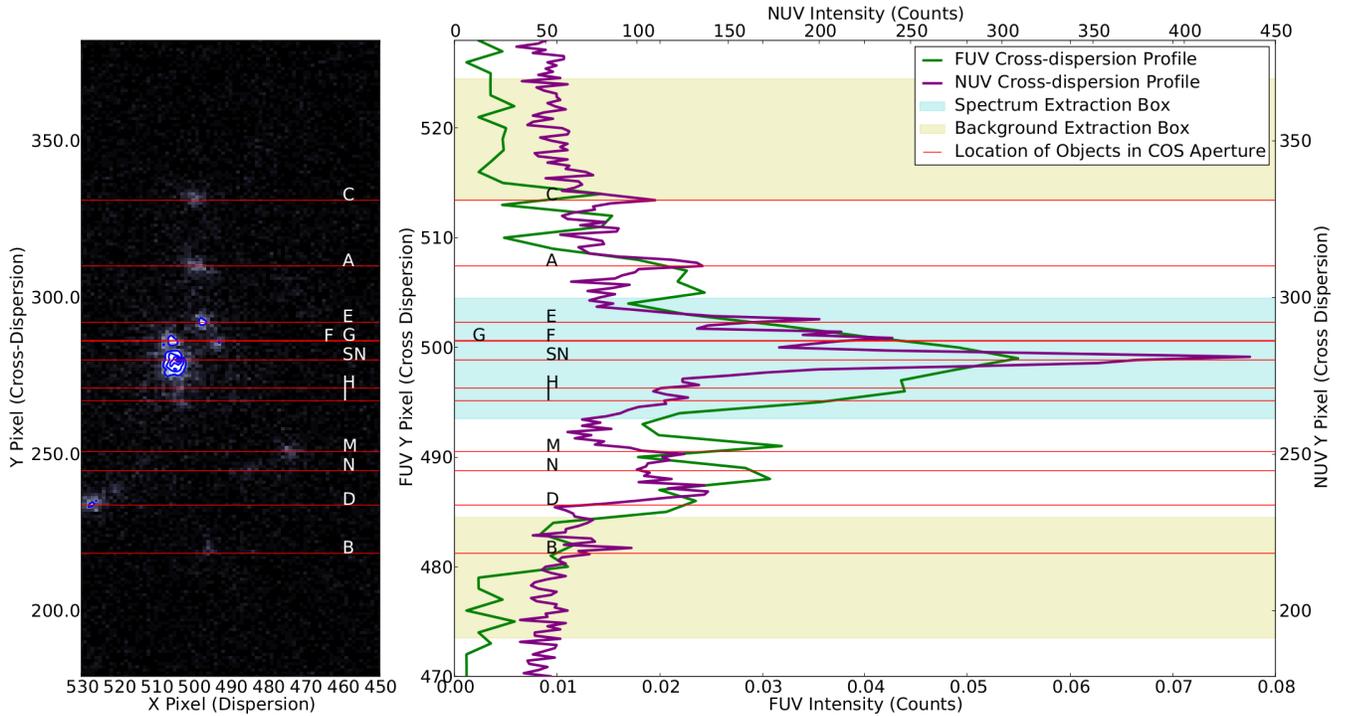}
\caption{Visit 07 TA image zoomed in on the target location (left). The cross-dispersion profiles of the FUV and NUV observations are in 
green and purple, respectively. The extraction box for the spectrum is shown in cyan and the extraction boxes for the background regions are in yellow. In both plots, the locations of the stars found in Figure \ref{fig1} are labeled and marked with horizontal red lines. It is apparent in the 
FUV cross-dispersion profile that Stars E--I have overlapping line-spread functions.}
\label{fig4}
\vspace{-0.2in}
\end{figure*}

The COS data consist of far-UV (FUV) and NUV observations taken over three visits (11 orbits).  The COS NUV observations use the G230L grating with central wavelengths 2635, 2950, and 3360~\AA. The 2635 and 3360~\AA\ central wavelengths were observed in a single visit using two FP-POS positions (dispersion direction dithers) for each central wavelength to correct fixed-pattern noise. The 2950~\AA\ central wavelength was observed in a second visit using four FP-POS positions. Although the COS NUV spectroscopic modes disperse light onto three stripes on the detector (similar to an echelle grating), the third stripe (Stripe C) of the G230L grating is contaminated with second-order light and the fluxes from this stripe are unreliable. We therefore exclude it from our analysis. The COS FUV data were obtained in the final visit using the G140L grating and the 1105~\AA\ central wavelength. The COS FUV mode typically places the spectrum across two detectors: Segment A and Segment B. However, when the 1105~\AA\ mode is used, Segment B must be turned off to avoid damaging the detector with zeroth-order light, so we present only Segment A data. The FUV data were taken using three FP POS positions.

Figure \ref{fig1}(d) displays the COS field of view.  Despite its large aperture diameter (2.5\arcsec), we chose COS for its superior resolution and low dark rate compared to STIS (see the COS Instrument Handbook).  Figure \ref{fig1} highlights that while SN 1993J is the brightest object in the aperture, a number of nearby sources are also UV bright. Even at this resolution their line-spread functions overlap and spatially contaminate the SN spectrum. Spectra were extracted using the COS calibration pipeline, CALCOS\footnote{http://www.stsci.edu/hst/cos/pipeline/ .}.  The default extraction parameters were modified to best isolate the SN 1993J flux from other objects in this crowded field.  We stress that even with $HST$'s superior angular resolution, isolating SN 1993J proved difficult owing to both the faint NUV flux and relatively low FUV cross-dispersion resolutions (see Fig. \ref{fig4}). 

\begin{deluxetable*}{l c c c c} 
\tablewidth{0pt}
\tabletypesize{\normalsize}
\tablecaption{Customized Spectral Extraction Parameters\tablenotemark{1}\label{tab5}}
% for the FUV and NUV 1D spectral extraction 
\tablecolumns{4}
\tablehead{
\multirow{4}*{Mode} & \colhead{93J (F,G,H,I)} & \colhead{93J (F,G,H,I)} & \colhead{Background} & \colhead{Background} \\
\colhead{} & \colhead{extraction} & \colhead{extraction} & \colhead{extraction} & \colhead{extraction box} \\
\colhead{} & \colhead{box center} & \colhead{box size (centered} & \colhead{location offsets} & \colhead{height (centered}\\
\colhead{} & \colhead{Seg A/ Seg B} & \colhead{on Col 1)} & \colhead{from spectrum} &  \colhead{on Col 3)}
}
\startdata
G140L/1105 & 496.24 / -- &11 &20 &11\\
G230L/2950 & 176.04 / 276.64 &39 &70 &39\\
G230L/2635 & 177.82 / 278.19  &39 &70 &39\\
G230L/3360 & 179.86 / 280.07 &39 &70 &39
\enddata
\tablenotetext{1}{All units in pixels.}
\end{deluxetable*}

\subsection{Defining the Extraction Box}

While the COS low-resolution mode offers superior spectral resolution and lower background rates compared to similar STIS modes, its large 
slitless aperture inevitably integrates light from sources close to the SN; see Fig. \ref{fig1}(d).  Figure \ref{fig4} plots the cross-dispersion (spatial) profile of all sources within the COS aperture, created by summing the two-dimensional images in the dispersion direction. The NUV and FUV spectra do not fall in a flat line across the detector. For this reason, a cross-dispersion profile created from all pixels in the dispersion direction will produce broad profiles which do not represent the true  width of the spectrum on the detector. To balance the spatial resolution with the signal-to-noise ratio (S/N), we sum over pixels 5000--7000 in the COS/FUV/G140L/1105 spectrogram and pixels 440--520 in the COS/NUV TA image.  This figure highlights that the default extraction boxes used by the {\it HST} pipeline (52 pixels for the NUV and 47 pixel for the FUV) integrate point-spread functions (PSFs) from Stars A through I in the resulting spectrum.  We therefore utilize the NUV TA image to optimize our extraction parameters. 

The COS NUV detector has the highest resolution of any instrument on {\it HST}, with 24 mas pixel$^{-1}$ and a 3-pixel resolution element in both the dispersion and cross-dispersion directions.  For the NUV spectra, a very narrow (10 pixel) extraction box can isolate the SN from the other stars, but this would result in significant flux loss from the wings.  Since the NUV counts are low to begin with [(1.5--3.5) $\times 10^{-4}$ counts s$^{-1}$], the additional flux from these wings proves important.  Furthermore, the FUV/G140L grating has a lower resolution of 80.3 mas pixel$^{-1}$ and a $6 \times 10$ pixel resolution element (dispersion by cross-dispersion).  Given the FUV resolution-element size and the S/N of the NUV data, we define the extraction box for both detectors as $\sim 1$ FUV cross-dispersion resolution element of 11 FUV pixels (39 pixels in the NUV) (see Fig. \ref{fig4} and Table \ref{tab5}).  Our resulting spectrum should therefore be considered a blend of the SN plus stars E through I, referred hereafter as ``SN+Stars E--I.'' In \S \ref{sec:disc} we will attempt to disentangle the different components by combining the photometry and spectroscopy with stellar models. 

The dark rate in the NUV detector does not vary with detector location.  To minimize contamination by other sources, we define the background region above Star C and below Star D (see Fig. \ref{fig4}).  Although the lower background region includes Star B, placing the window any lower in the NUV results in the overlapping of the background region for Stripe A with the spectrum for Stripe B.  The FUV detector dark rate does vary with detector location, but these same background windows receive the minimum contamination from other sources.  For consistency, we therefore define the same background regions in the FUV.  It should be noted that owing to the faint SN flux, the background count rate dominates the total count rate.  A visual inspection of the background at different extraction box heights confirms that the chosen background extraction box yields accurate background rates.  Table \ref{tab5} and Figure \ref{fig4} highlight the spectrum extraction box center and height, as well as the two background region extraction box centers and heights for each central wavelength.

\subsection{Extracting with a Nonstandard Box Height}

All extractions are performed using the COS calibration software CALCOS 2.20.1.  When not using the default extraction parameters, however, two additional details must be considered.  First, the flux calibration in CALCOS does not correct for the light which falls outside a nonstandard extraction box.  Second, the default centering of the extraction box assumes that the target is centered in the PSA.  The location of the PSA on the detector is identified with the wavelength calibration line lamp (wavecal) spectrum and a known offset between the wavecal aperture and the center of the PSA.  Since the default extraction box is typically large, the centering is only required to be accurate to within a few pixels.  The SN 1993J extraction requires a much smaller extraction box height, so the location of the extraction box center must be more precise.

The SN 1993J COS observations are very faint and blended, so they cannot be used to directly calculate either the aperture correction or extraction box center.  Instead, we utilize the spectrum of bright, well-characterized white dwarfs used to monitor the sensitivity of both the COS FUV and NUV detectors over time.  The bright and well-defined fluxes allow for an accurate measurement of the spectrum center and flux corrections for small extraction boxes.  Furthermore,  the large number of regular observations throughout COS operations allow us to measure the repeatability of these values.

\begin{figure}[h]
\includegraphics[width=0.5\textwidth]{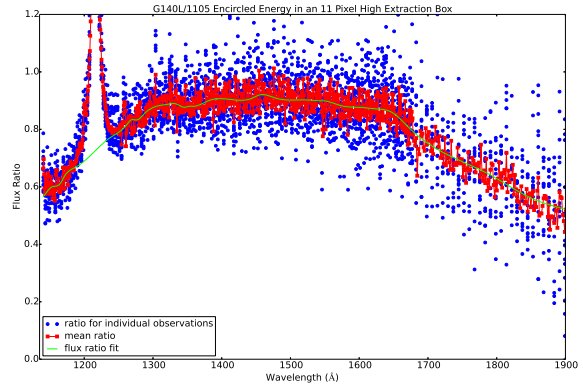}
\caption{The ratio of the flux in an 11-pixel-high extraction box with nonstandard background regions to the flux in a default extraction as a function of wavelength for G140L/1105 sensitivity monitoring observations. The ratio for individual observations is in blue, the mean of each observation in red, and a cubic spline fit to the mean in green. The mean data values are linearly interpolated in the range 1185--1250~\AA\ to avoid fitting the Ly$\alpha$ airglow line.}
\label{fig5}
\vspace{0.1in}
\end{figure}

\begin{deluxetable*}{l l l l} 
\tablewidth{0pt}
\tabletypesize{\normalsize}
\tablecaption{Wavelength Ranges Excluded from Final COS Spectrum \label{tab6}}
\tablecolumns{4}
\tablehead{
\colhead{Mode} & \colhead{Segment/Stripe} & \colhead{Flagged (\AA)} & \colhead{Reason} \\
}
\startdata
G140L/1105 &FUVA &1198--1202.5 &Geocoronal N~I\\
G140L/1105 &FUVA &1210.5--1220.5 &Geocoronal Ly$\alpha$\\
G140L/1105 &FUVA &1300.5--1307.4 &Geocoronal O~I\\
G140L/1105 &FUVA &1354.3--1359 &Geocoronal O~I\\
G140L/1105 &FUVA &$\lambda >1665$ &Low sensitivity\\
G230L/2635 &NUVA &$\lambda <1665$ &Low sensitivity\\
G230L/2950 &NUVB &$\lambda >3000$ &Low sensitivity, second-order light\\
G230L/3360 &NUVB &$\lambda > 3000$ &Low sensitivity, second-order light
\enddata
\end{deluxetable*}

The CALCOS \emph{x1d} task was run on 14 G140L/1105 observations of WD0947+857 using the extraction parameters detailed in Table 
\ref{tab5} for SN 1993J. Figure \ref{fig5} plots the ratio of the FUV flux-calibrated spectrum with these custom extraction  parameters to the CALSPEC model spectrum\footnote{http://www.stsci.edu/hst/observatory/cdbs/calspec.html .} for each observation. A cubic spline is fit to the mean value of the observations as a function of wavelength. The uncertainty of the fit at each wavelength is calculated by fitting a cubic spline to the standard deviation of the observations around the fit.  The fit to the flux ratio is used to empirically correct the flux of the extracted spectrum of SN~1993J.  A similar procedure is used for each stripe of the NUV detector.

By default, the \emph{x1d} task extracts the target spectrum from the nominal spectrum location combined with the location of the aperture on the detector. However, it is possible to ask CALCOS to find the spectrum location. This option is used on the white-dwarf spectra to determine if a systematic offset exists between the nominal spectrum location and the actual spectrum location. Offsets of 2 pixels and $-3$ pixels are found for the G140L/1105 and G230L/2635 modes, respectively.

\subsection{Generating the COS Spectrum}

\begin{figure}
\begin{center}
\hspace*{-0.4in}\includegraphics[width = 0.55\textwidth]{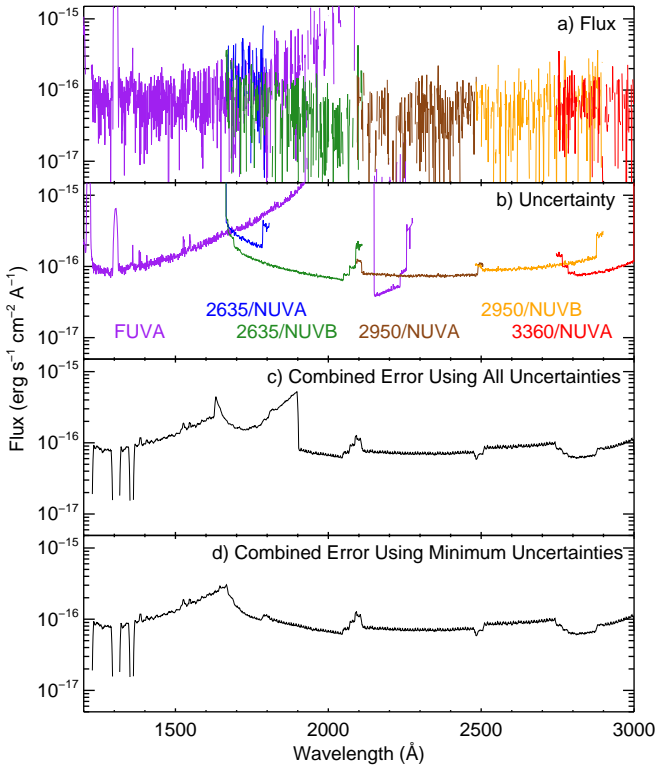}
\caption{Plot of spectra extracted for each central wavelength.  (a) Individual flux spectra.  (b) Individual uncertainties.  (c) Resulting error bars when flux spectra combined using the \emph{splice} task in the \emph{PyRAF STSDAS} package.  (d) Resulting error bars when final spectrum built by choosing only the minimum error and associated flux at a given wavelength (i.e., no coadding of spectra).  The effects of these methods are discussed in \S \ref{sec:chi}.
}
\label{fig6}
\end{center}
\end{figure}

Each observation of SN 1993J is extracted using the parameters listed in Table \ref{tab5} and the offset location found above. 
The different FP-POS positions of each central wavelength are combined using the \emph{fpavg} task in CALCOS to create one spectrum for each 
central wavelength. The flux for each central wavelength is then corrected to an infinite aperture and the errors are scaled appropriately.  Figure \ref{fig6} plots the flux and observational uncertainty for each central wavelength.  Spectra from each central wavelength are combined into a single spectrum using the \emph{splice} task in the \emph{PyRAF STSDAS} package.  Prior to combining the central wavelengths, regions of low sensitivity and geocoronal airglow are flagged as poor quality to avoid contaminating the final spectrum (see Table \ref{tab6} for a list of flagged regions).

\begin{figure}[t]
\vspace{0.1in}
\hspace{-0.3in}\includegraphics[width=0.5\textwidth]{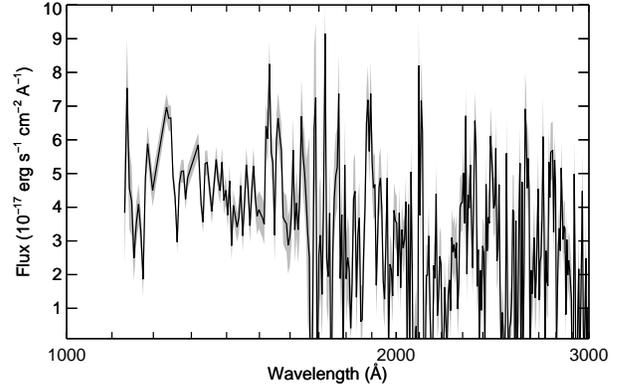}
\caption{The COS FUV and NUV observations combined into a single spectrum with regions of poor data quality excluded. The spectrum is in black with uncertainties in grey.  The spectrum has been boxcar smoothed by 70 pixels.}
\label{fig7}
\end{figure}

Figure \ref{fig6} shows, however, that the estimates of the observational uncertainties associated with the combined spectrum are not well behaved at wavelengths where the FUV and NUV spectra overlap.  A detailed analysis in Section \ref{sec:chi}, below, further shows that over large regions of the spectrum, observational uncertainties are overestimated by varying factors depending on which central-wavelength grating is used.  Instead, we find that better-behaved observational uncertainties can be obtained by not coadding spectra at wavelengths where the FUV and NUV spectra overlap.  We choose to extract only the minimum observational uncertainty and associated flux at these wavelengths.  The excluded regions are listed in Table \ref{tab6}.  Figure \ref{fig6} also plots the resulting uncertainties for this technique.  In Section \ref{sec:chi}, we further show that these results are substantially more uniform over the entire spectral range, and the overall distribution is very close to normally distributed.

%%Line Profiles
\begin{figure*}[t]
\begin{center}
\epsscale{1.}
\plotone{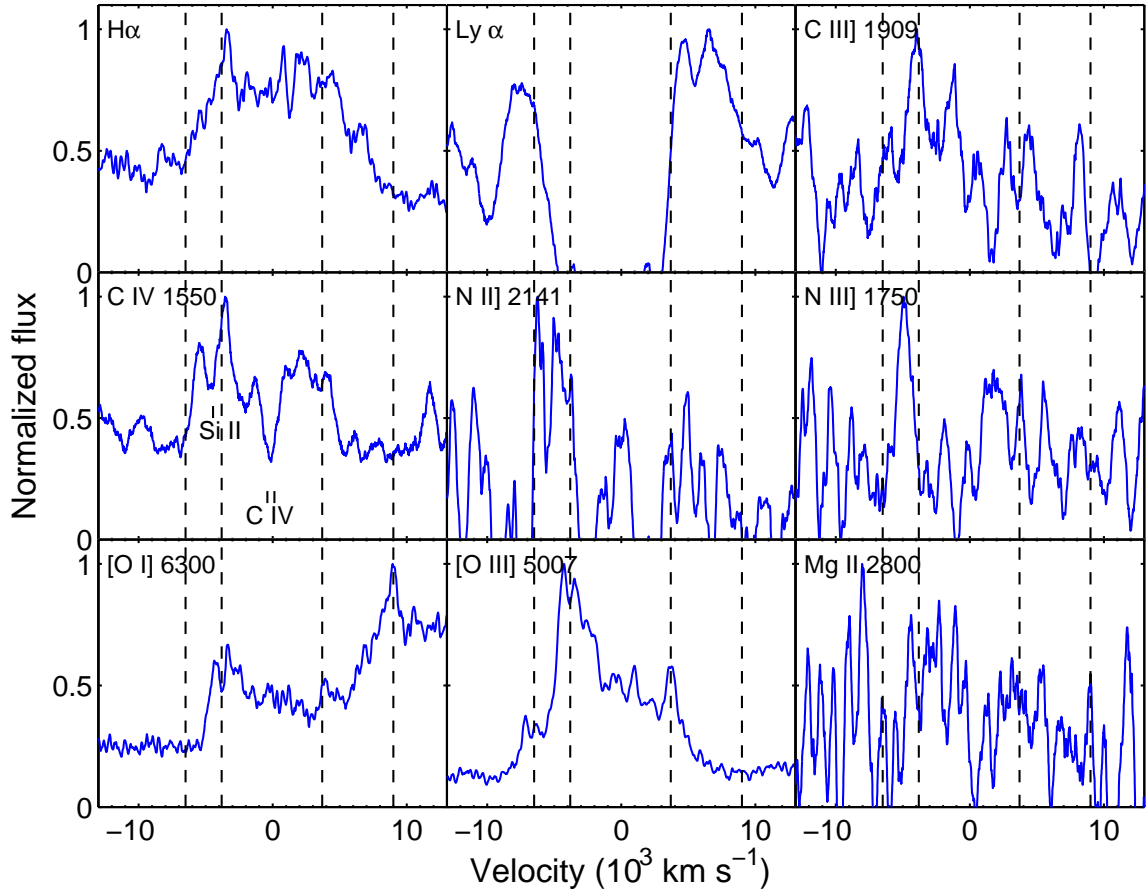}
\caption{Compilation of UV and optical emission-line profiles. The flux of each line is normalized within the velocity range $\pm$10,000~\kms. The positions of the Si~II $\lambda$1526 and C~IV interstellar absorptions are marked in the C~IV panel. Dashed lines indicate [O~III] line peaks at $\sim -3800$~\kms\ and +3700~\kms.  Dashed lines also mark the constraints on the ejecta velocity given by the H$\alpha$ line at $\sim -6000$~\kms\ and $\sim +9000$~\kms.}
\label{fig8}
\end{center}
\end{figure*}

Figure \ref{fig7} plots the final COS spectrum, which should be considered a sum of flux from the SN+Stars E--I.  These are the same sources that must have been included in the Keck/LRIS spectrum presented by \citet{maund04}.  Unlabeled fainter stars are apparent in the WFC3/F336W image in Figure \ref{fig1}, but are not detected in the COS NUV image.  We thus conclude that these stars do not contribute significantly to the extracted COS spectrum.

\subsection{Spectral Line Identifications and Profiles}

Figure \ref{fig8} shows a compilation of the most prominent UV lines in the COS spectrum.  We also include several optical lines from the 2013 February 17 Keck spectrum (see \S \ref{sec:keck}). In this latter spectrum the H$\alpha$~and [O~III] $\lambda\lambda$4959, 5007 emission lines are strongest.  Even the blended [O~III] doublet shows considerable structure, which reflects real fluctuations in the emissivity as a function of the velocity.  The reality of these fluctuations is demonstrated by the same two peaks at $-3472$~\kms\ and $-4272$~\kms\ in both the $\lambda$4959 and $\lambda$5007 components of the [O~III] doublet.  The [O~III] line shows an additional asymmetry favoring the blue side, which agrees with observations of SN 1993J by \citet{milisavljevic12}.
%, as well as some of the weaker lines from the day 2579--2585 STIS spectrum

Several other lines show structure similar to that of [O~III]. In particular, the peak at $\sim -3800$~\kms\ is clearly present in the H$\alpha$, Ly$\alpha$, [O~I], C~III], and C~IV lines. Note the interstellar absorption of the Si~II $\lambda$1526 line, which distorts the C~IV peak. The N~II], N~III], and Mg~II lines have a lower S/N, and for these lines the $\sim -3800$~\kms\ peak is the most significant extension above the noise.  Dashed lines in Figure \ref{fig8} highlight the strongest peaks at both $\sim -3800$~\kms\ and $\sim +3700$~\kms. These velocities also correspond to the flat part of the box-like line profiles previously identified by \citet{fransson05}.  

The asymmetry is less pronounced in the H$\alpha$ line compared to the [O~III] line, although the fluctuations are present.  These lines likely arise from different components, with the H$\alpha$ line coming mainly from the dense, cool shell behind the reverse shock, while the [O~III] and other high-ionization lines come from the ionized outer parts of the unshocked ejecta (e.g., \citealt{chevalier03}).  The full extension of the lines, marking the maximum ejecta velocity, is somewhat difficult to determine because of the gradual transition to the continuum. The red wing of the H$\alpha$ line extends to $\sim +9000$~\kms, although the maximum velocity may be influenced by the [N~II] $\lambda$6583 line. The blue wing is blended with the [O~I] $\lambda\lambda$6300, 6364 lines, which allows us to constrain an ejecta velocity of at least $\sim 6500$~\kms\ on this side.  Figure \ref{fig8} also highlights these velocities with dashed lines.  Compared to high-ionization lines (i.e., the C~IV and [O III]), H$\alpha$ has a larger width on both the red and blue sides, with a maximum velocity of 5000--6000~\kms.  As discussed above, these velocities may arise from different regions where these lines are formed. In particular, the high-ionization lines are likely to be formed by photoionization in the processed gas in the unshocked core, and are consequently expected to have lower velocity that the H$\alpha$ line.

\section{Analysis}
\label{sec:disc}

\citet{maund04} detected a number of absorption lines in the NUV portion of their Keck/LRIS spectrum of SN 1993J.  Models indicate that these lines are most consistent with those of a B2~Ia star at the position of SN 1993J.  \citet{maund09} later found a NUV excess in an {\it HST} SED consistent with the combined flux from a B2~Ia star and SN~1993J, although the SN still dominated the SED.  \citet{maund09} predicted that by 2012, the SN would fade sufficiently to directly measure the properties of the putative remaining hot B companion star.  Here we disentangle the contributing sources in the 2012 COS spectroscopy. 

\subsection{Defining a Supernova Template}
\label{sec:template}

Since the SN and putative companion are spatially coincident, we must first define a template representative of the SN contribution to the overall spectrum.  Prior to 2012, the most recent UV observations of SN 1993J were taken using the Space Telescope Imaging Spectrograph (STIS) in April 2000 with the G140L and G230L gratings (GO-8243; PI R. Kirshner; \citealt{fransson05}).  This spectrum serves as a useful reference template; we can assume that by this epoch the SN was quite evolved, and we expect little change in the spectrum besides continued fading.  The validity of this assumption is highlighted by the slow UV spectral evolution in 1995--2000 shown by \citet[][their Figure 1]{fransson05} and the slow optical spectral evolution in 2000---2009 shown by \citet[][their Figures 10 and 11]{milisavljevic12}.  While some line strengths do vary, the underlying continuum shape and line profiles of the strongest emission lines do not change significantly, which is most important for this study.

%%UV Spec Comparison
\begin{figure}[t]
\begin{center}
\epsscale{1.}
\hspace{-0.1in}\includegraphics[width=0.48\textwidth]{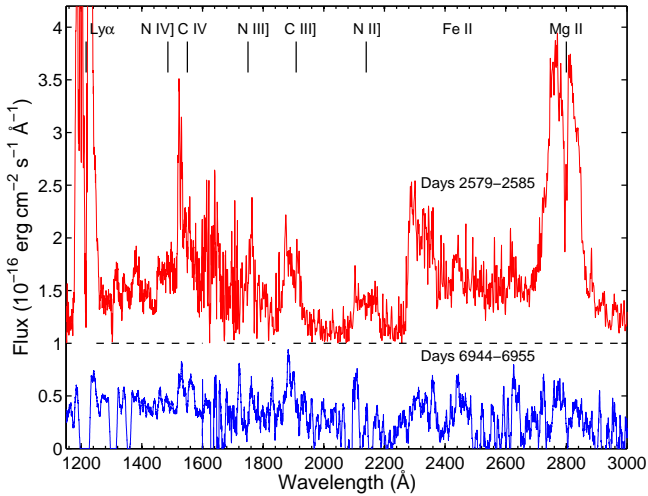}
%\plotone{fig9.eps}
\caption{UV spectrum of SN 1993J from days 2579--2585 and 6944--6955. The most prominent emission lines have been marked. The top spectrum has been shifted up by $1\times10^{-16}$~erg~cm$^{-2}$~s$^{-1}$~\AA$^{-1}$~for clarity.}
\label{fig9}
\end{center}
\vspace{-0.2in}
\end{figure}

For the 2000 STIS observations, the SN flux dominated over all neighboring stars, so contamination is not an issue.  We therefore use the default pipeline product, just as \citet{fransson05} did.  These observations are again combined using the \emph{splice} task. Like the COS spectrum, wavelengths of low sensitivity and geocoronal airglow are flagged prior to the combination. A list of excluded wavelength ranges can be found in Table \ref{tab7}. 

%\subsubsection{Comparing the 2000 and 2012 Spectra}
%\label{sec:model}

Figure \ref{fig9} compares the 2000 STIS and 2012 COS spectra.  Although the S/N is considerably lower in the 2012 spectrum, most of the lines identified in 2000 are also observed in 2012.  As in 2000, the strongest line is Ly$\alpha$, but it is severely affected by the strong geocoronal emission line, the geocoronal subtraction, and the damping wings of the interstellar Ly$\alpha$~absorption.  We therefore do not include this line in our analysis, although we do note that it appears asymmetric with a strong blue wing.  Both the C~IV $\lambda\lambda$1548, 1551 and C~III] $\lambda$1909 lines are still strong in the 2012 spectrum.  The N~II] $\lambda\lambda$2140,2144 doublet is weaker, but still above the noise. The N~III] $\lambda\lambda$1747--1754 multiplet is in a noisy region of the spectrum, but the blue peak is above the noise. In \citet{fransson05}, the feature at 2335~\AA\ was identified with C~II] $\lambda\lambda$2323, 2328, O~III] $\lambda\lambda$2321, 2331, and Si~II] $\lambda$2335, and it is also well defined in the 2012 spectrum. The peaks at $\sim 2460$~\AA~and $\sim 2640$~\AA~are probably produced by Fe~II resonance lines. The most dramatic change is the strength of the Mg~II $\lambda\lambda$2796, 2803 doublet, which was the strongest UV line in 2000 but is only barely above the noise in 2012.

%%Best Fit w/ STIS Spectrum
\begin{figure}[b]
\begin{center}
%\epsscale{1.2}
\hspace{-0.3in}\includegraphics[width=0.5\textwidth]{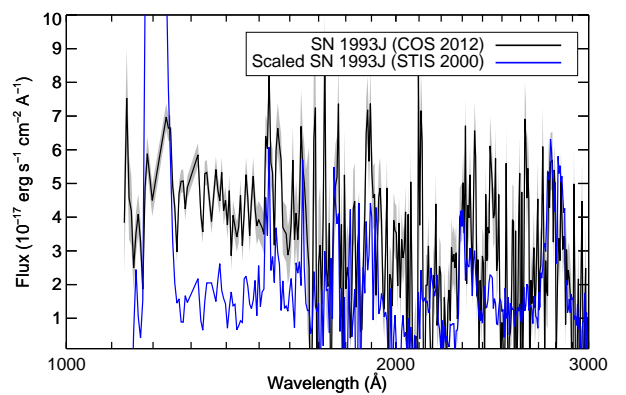}
%\plotone{fig10b.eps}
\caption{The COS UV spectrum fit with only the year-2000 STIS spectrum scaled by a factor of 0.31.  This scaled STIS fails most significantly in fitting the rising FUV continuum.  Subsequent figures describe additional components.
}
\label{fig10}
\end{center}
\vspace{-0.15in}
\end{figure}

\subsection{Modeling the COS Spectrum}
\label{sec:nuvspectrum}

%\subsection{Modeled vs. Observed NUV Flux}
%\label{sec:nuvflux}

Figure \ref{fig7} plots the combined COS NUV and FUV spectrum.  We again stress that the spectrum should be considered a sum of flux from the SN+Stars E--I (see \S \ref{sec:cos}).  Since the optical wavelengths are dominated by the fading SN itself, we aim here to fit the combined NUV and FUV spectrum (i.e., $\lambda < 3000$~\AA).  This wavelength range is not sensitive to the same stellar absorption lines observed by \citet{maund04} at $> 3600$~\AA\ (see spectral models below).  To model the data, we consider all possible components with the intent of minimizing $\chi^2$ (see \S\ref{sec:chi}).

%%Best Fit w/ Field Stars
\begin{figure*}[t]
%\rotate
\begin{center}
\epsscale{0.95}
\plotone{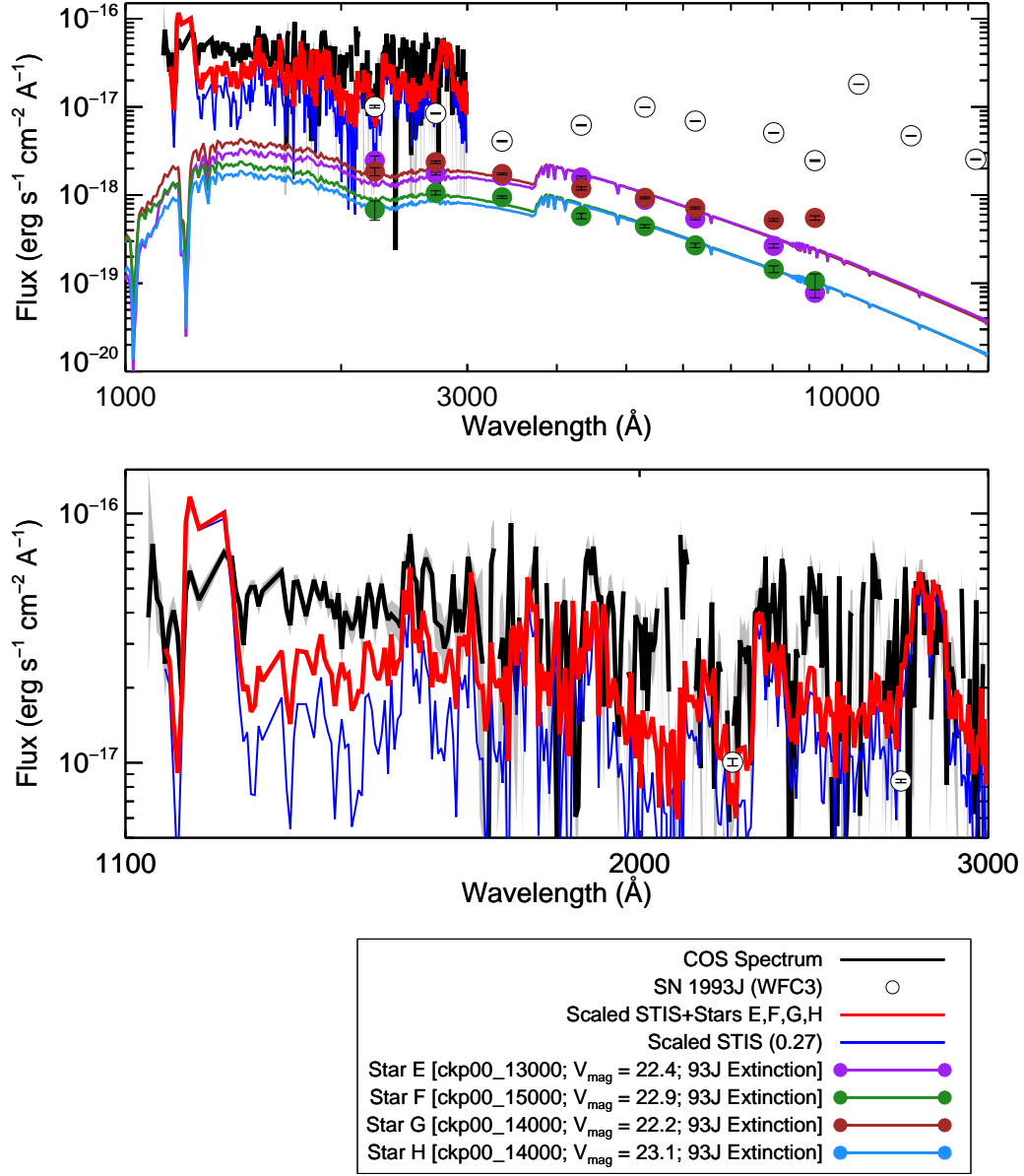}
\caption{The COS UV spectrum is better fit by including stellar models for Stars E, F, G, and H along with the year-2000 STIS spectrum scaled by a factor of 0.28.  Nonetheless, the FUV continuum excess remains unaccounted for.  (Top) The entire {\it HST} dataset, including COS spectra and WFC3 photometry.  The longer-wavelength photometry constrains the stellar models used for Stars E--H.  The excess flux for SN 1993J at these longer wavelengths is attributed almost entirely to the SN component and not the putative companion; we therefore do not fit these data.  (Bottom) A closer look at the UV spectrum and best-fit model.%  These components result in $\chi^2=0.60$.
}
\label{fig11}
\end{center}
\end{figure*}

{\bf Fading SN:}  As described in \S \ref{sec:template}, we adopt the UV spectrum obtained with $HST$/STIS in 2000 \citep{fransson05} as the template of the fading SN 1993J.  We use the IDL {\tt MPFIT} nonlinear least-squares curve fitting function \citep{markwardt09} to find the best fit by varying a single multiplicative scale factor across the entire spectrum.  Figure \ref{fig10} shows that a best fit can be achieved with a scale factor of 0.31 times the original STIS flux in 2000.  At longer wavelengths (the NUV), the continuum levels match well.  The lines, however, vary in their strength and do not appear to be well represented by a single scale factor.  At shorter wavelengths, a FUV continuum excess exists which can be best fit by additional components described below. 

%\subsection{The final STIS spectrum}
\begin{deluxetable}{l l l}
\tablewidth{0pt}
\tabletypesize{\normalsize}
\tablecaption{Wavelength Ranges Excluded from Final STIS Spectrum \label{tab7}}
\tablecolumns{3}
\tablehead{
\colhead{Mode}	& \colhead{Flagged (\AA)}	& \colhead{Reason} \\
}
\startdata
G140L	&$\lambda <1141$	&Low sensitivity \\
G140L	&1206--1220	&Geocoronal Ly$\alpha$ \\
G230L	&$\lambda <1700$	&Low sensitivity 
\enddata
\end{deluxetable}

%%Best Fit w/ B2 Star
\begin{figure*}[t]
%\rotate
\begin{center}
\epsscale{0.95}
\plotone{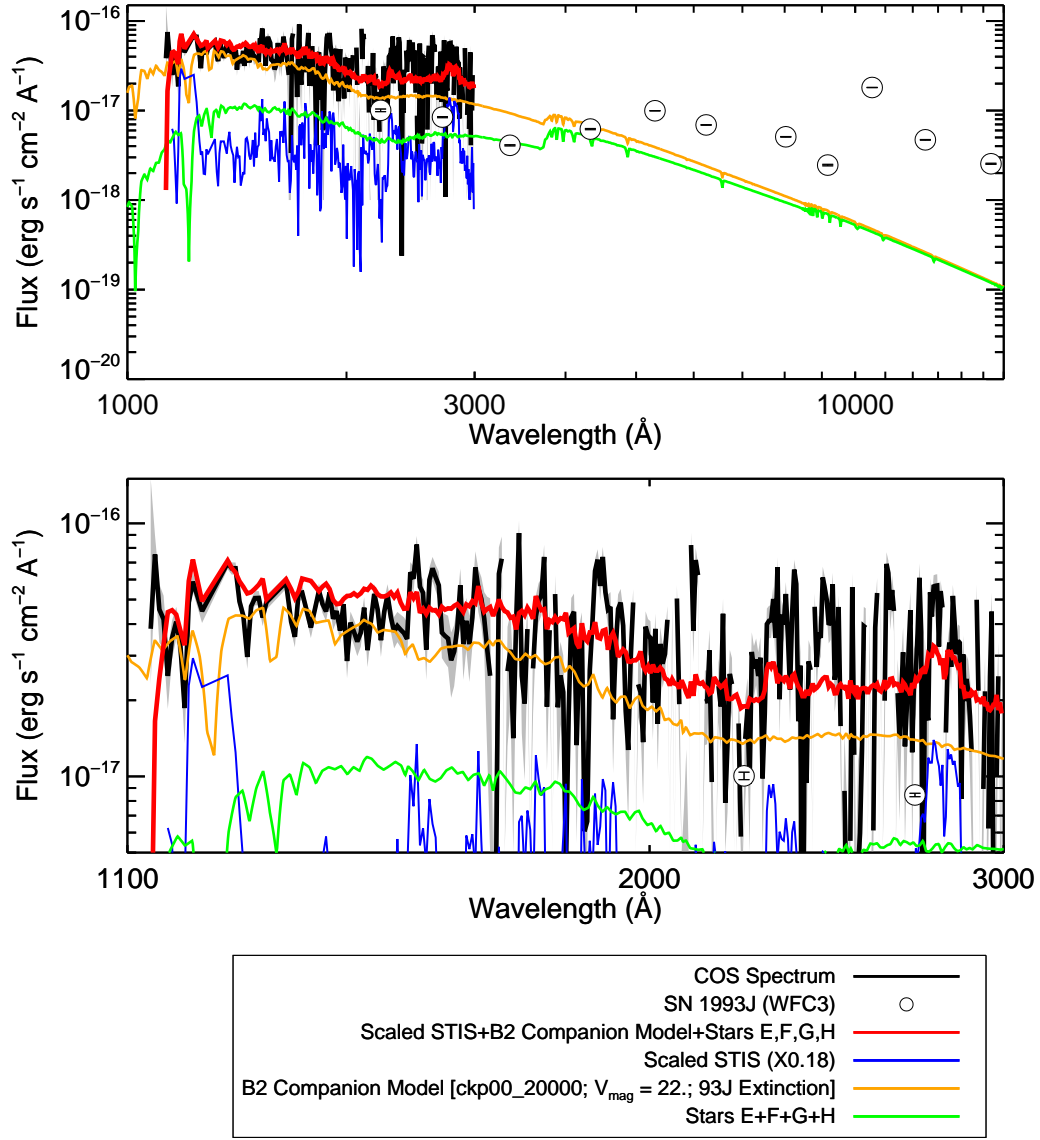}
\caption{Same as Figure \ref{fig11}, except with an additional component representing a B2~Ia $T =$ 20,000~K stellar model normalized to $V=22$ mag (orange).  While the overall fit is better than in Figure \ref{fig11}, the B2~Ia model is brighter than the WFC3 photometry for SN 1993J. 
}
\label{fig12}
\end{center}
\end{figure*}

%%Best Fit w/ Modified B2 Star
\begin{figure*}[t]
\begin{center}
\epsscale{0.95}
\plotone{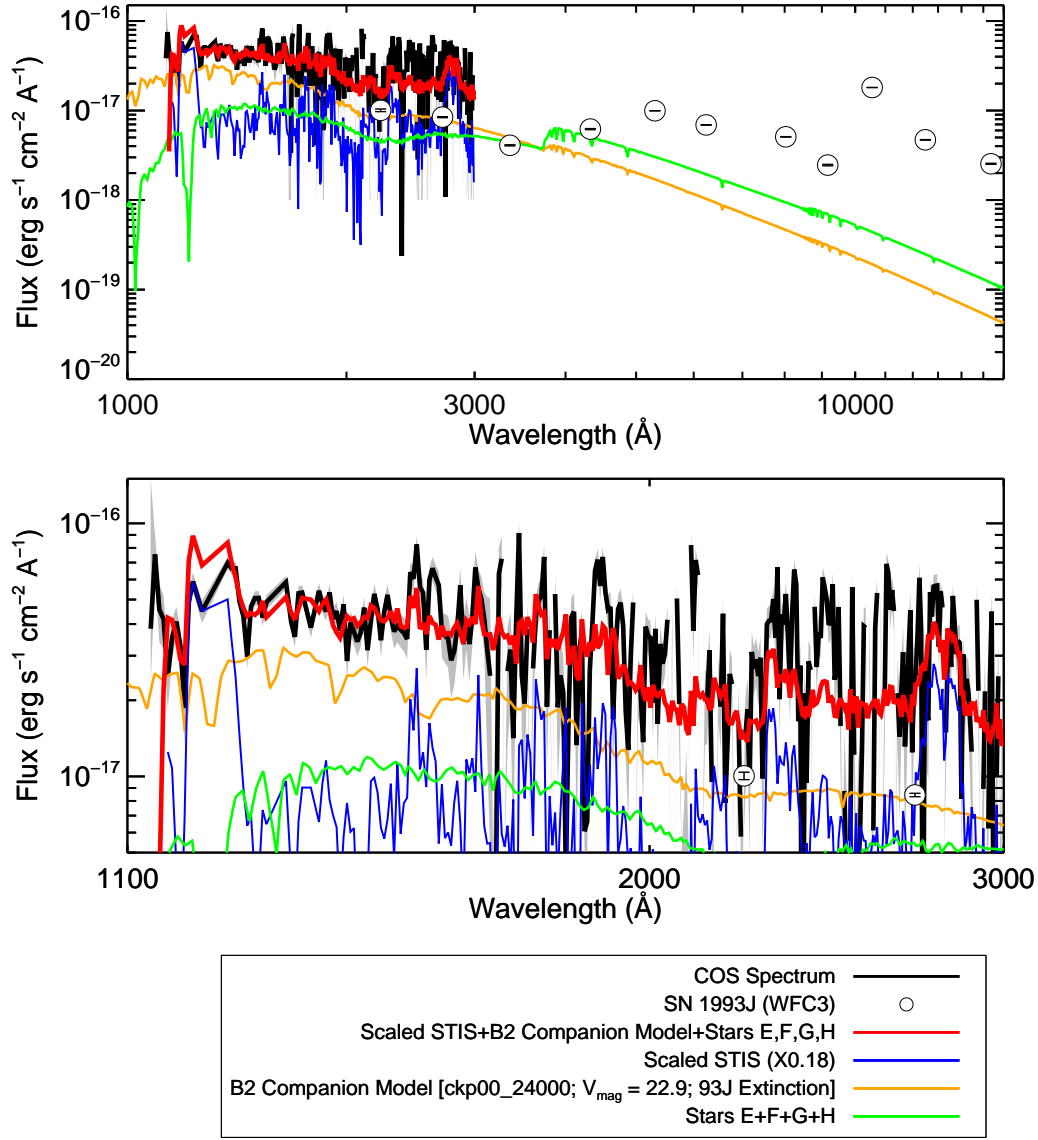}
\caption{Same as Figure \ref{fig12}, except with a hotter B1 $T =$ 24,000~K stellar model normalized to $V=22.9$ mag (orange).  The value of $\chi^2$~was minimized for this model while remaining consistent with the WFC3 photometry for SN 1993J.
}
\label{fig13}
\end{center}
\end{figure*}

%%Minimizing Chi Squared
\begin{figure}[t]
%\rotate
\begin{center}
%\epsscale{0.95}
\hspace{-0.3in}\includegraphics[width=0.5\textwidth]{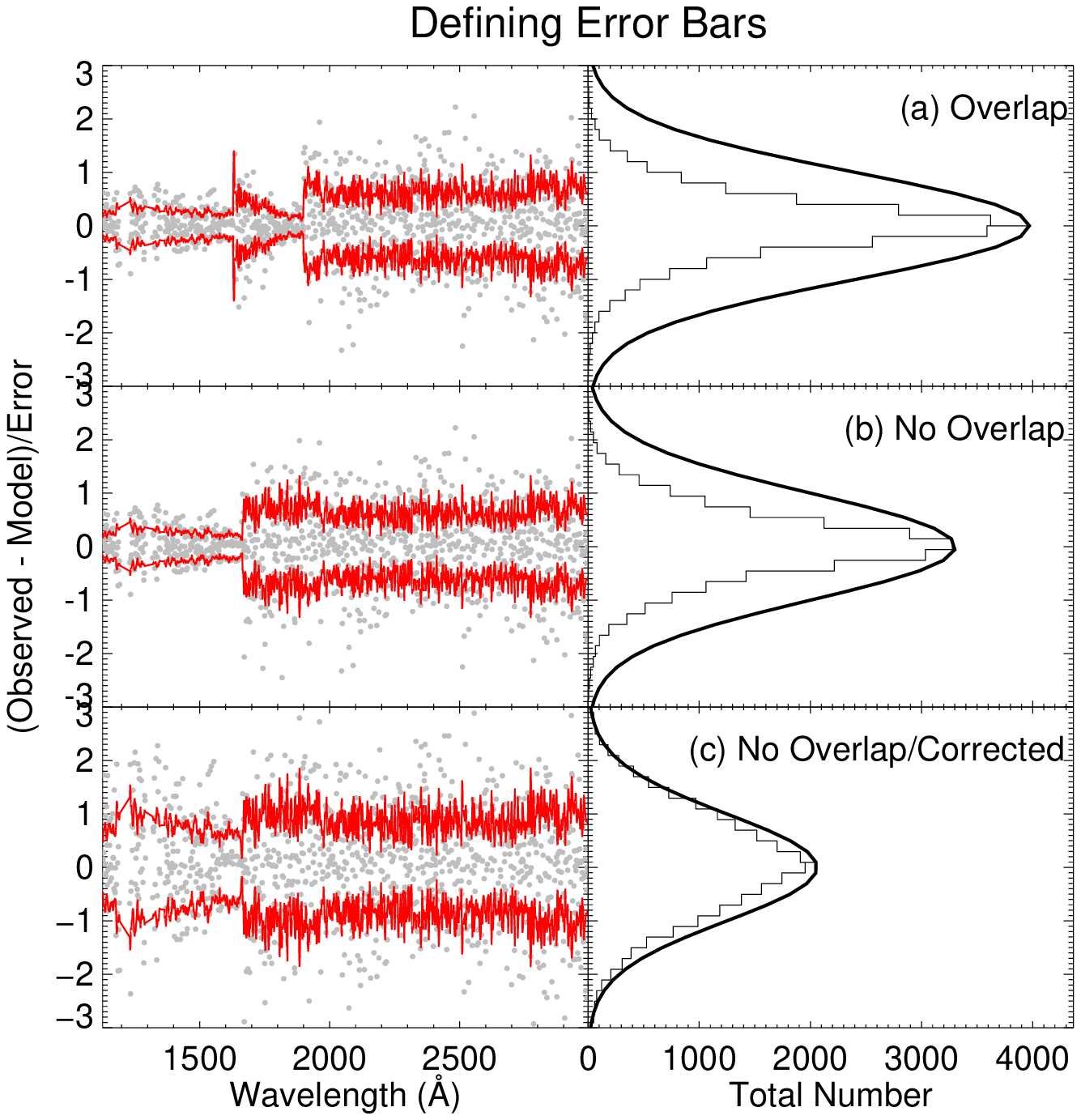}
%\plotone{fig14.eps}
\vspace{-0.2in}
\caption{For each method used to generate the final COS spectrum, the errors are represented by the distribution of $\chi = ({\rm Observed}(i) - {\rm Model}(i))/{\rm Uncertainty}(i)$, for which a normal distribution should be characterized by a Gaussian with a standard deviation equal to 1.  (left) $\chi$~vs. wavelength.  (right) Histogram of the $\chi$~distribution, overplotted with a Gaussian with a standard deviation equal to 1.  (a) Results from using the \emph{splice} task in the \emph{PyRAF STSDAS} package. (b) Results when selecting only the minimum error and associated flux at a given wavelength (i.e., no coadding). (c) Same as (b), but the errors are manually corrected to achieve a normal distribution with a standard deviation close to 1.  Ultimately, (c) illustrates the method used to generate the COS spectrum used for our modeling in \S \ref{sec:nuvspectrum}.
}
\label{fig14}
\end{center}
\end{figure}

{\bf Stars E--H:}  Figure \ref{fig11} plots the WFC3 photometry for SN 1993J and Stars E--H (Star I was too faint to be detected by Dolphot).  The COS spectrum blends Stars E--I together with the SN, but the WFC3 photometry isolates the flux from the individual stars.  For this reason, the SN 1993J photometry does not match the spectrum.  The combined flux from all the stars, however, is consistent with the spectrum to within the uncertainties.

We fit Stars E--H using Castelli and Kurucz models \citep{castelli04} convolved with the WFC3 throughput for each filter using IRAF's synphot.calcspec package.  For simplicity, we assume the same extinction as the SN (see below).  The model chosen for each star is the model for which $\chi^2$~is minimized.  These stellar models provide reliable estimates of the FUV fluxes despite a lack of photometry at these wavelengths.  Figure \ref{fig11} shows the resulting best fits.  We then rescale the STIS spectrum to account for the remaining flux.  Combined with the flux from Stars E--H, a STIS scale factor of  $\sim 0.28$ results in a best fit, but there is still an FUV excess implying the existence of another component. 
%0.60

{\bf Additional B2 Star:}  The contribution of FUV flux from Stars E--H is minimal, and even with the addition of these stars, a FUV excess is still apparent.  We therefore consider the presence of a binary companion, previously suggested to be a B2~Ia star \citep{maund04,maund09}.  We generate a B2~Ia spectrum using the Castelli and Kurucz model in IRAF's synphot.calcspec package.  We use a 20,000 K star with a surface gravity log($g$/cm-s$^{-2}$) = 3.5 and metallicity [M/H] = 0.0.  Following \citet{maund04}, we renormalize to the observed $V$-band magnitude ($V_{\rm Vegamag} = 22.0$ accounting for an extinction of $A_V = 0.52$ mag to SN 1993J).  Again, we use the IDL {\tt MPFIT} function to find the best fit by varying only the multiplicative scale factor for the STIS spectrum.  

Figure \ref{fig12} shows an improved fit with a STIS scale factor of $\sim 0.12$.  This model, however, is not ideal.  The combined flux from the scaled STIS spectrum and B-star model (orange) should be consistent with the observed WFC3 photometry, but in this case the B-star model is too bright.  To optimize our fit, we consider other stellar types and luminosity classes ranging from 10,000~K to 31,000~K, as well as normalization magnitudes (although we do not vary the metallicity or surface gravity values).  We minimize $\chi^2$~(e.g., maximize the probability function) with a temperature of 24,000 K renormalized to $V_{\rm Vegamag} = 22.9$ mag (see Fig. \ref{fig13}).  Given the noise associated with the spectrum, the data are somewhat degenerate with several models that include B stars in the range 19,000--29,000~K (B3--B1).  We thus infer the presence of a spectral component consistent with continuum from a hot B star.

\subsection{Minimizing Reduced $\chi^2$}
\label{sec:chi}

To effectively utilize the reduced $\chi^2$~requires a detailed understanding of the observational uncertainties.  CALCOS, the COS calibration pipeline, generates both a flux-calibrated one-dimensional spectrum and an associated uncertainty.  This uncertainty is calculated using the formula given by \citet{gehrels86} (see COS Data Handbook; \citealt{massa13}). Gehrels' formula, however, offers only an upper limit to the uncertainty regardless of the number of source counts (for large counts the formula approaches the typical $\sqrt{n}$ error approximation).

Figure \ref{fig6} shows two methods used for generating our spectrum of SN 1993J.  In the first, we combine all spectra from each central wavelength into a single spectrum.  In the second, we do not coadd spectra at wavelengths where the FUV and NUV spectra overlap.  We choose to extract only the minimum observational uncertainty and associated flux at these wavelengths.  For each case, the residual errors can be characterized by the distribution of $\chi = ({\rm Observed}(i) - {\rm Model}(i))/{\rm Uncertainty}(i)$, for which a normal distribution should be characterized by a Gaussian with a standard deviation equal to 1.  For the two techniques described here, Figures \ref{fig14}(a) and  \ref{fig14}(b) show the values $\chi$ as a function of wavelength and the resulting histogram.

Figure \ref{fig14}a varies significantly, while Figure \ref{fig14}b shows a more normal distribution.  Still, the uncertainties appear too large in the FUV (i.e., $\chi<1$).  To compensate, we decrease the size of all uncertainties in spectral elements with a wavelength shorter than 1660 \AA~by a factor of 2.9, and all spectral elements with a wavelength longer than 1660 \AA~by a factor of 1.4.  Figure \ref{fig14}c shows the results, which is nearly a normal distribution.  We adopt these uncertainties throughout the modeling in this paper.

\begin{figure}[t]
\vspace{0.05in}
\begin{center}
\epsscale{0.95}
\hspace{-0.3in}\includegraphics[width=0.5\textwidth]{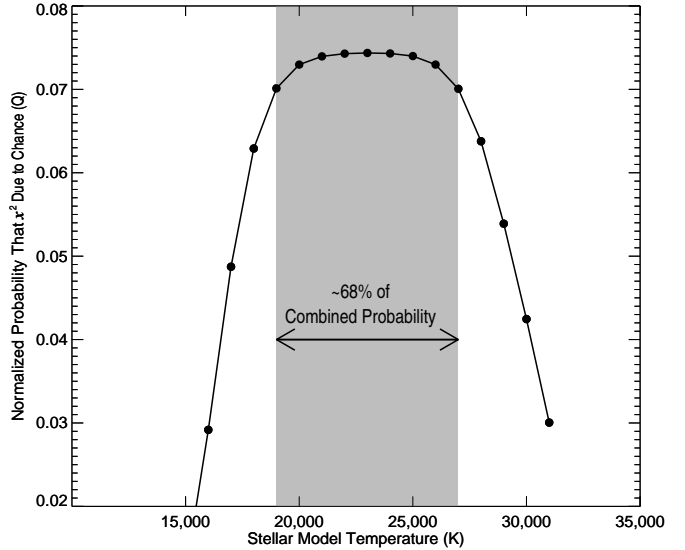}
%\plotone{fig15.eps}
\caption{The probability ($Q$) that the best-fitting model $\chi^2$~value is due to chance, normalized relative to the sum of the resulting probability distribution.  Models summing to 68\%~of the combined probability range from 19,000~K to 27,000~K.
}
\label{fig15}
\end{center}
\end{figure}

Despite the improved accuracy of the uncertainties, the resulting reduced $\chi^2$~values cannot be considered meaningful on an absolute scale because the degrees of freedom are not well defined.  We therefore calculate the probability ($Q$) that the resulting $\chi^2$~value is due to chance and normalize relative to the sum of the resulting probability distribution (see Figure \ref{fig15}).  We consider models summing to 68\%~of the combined probability (e.g., 19,000--27,000~K) to be the likely range of models that can fit the COS spectrum.

%The probability, $Q$, is $\propto e^\frac{\chi^2}{a}$, where $a$~is a scale factor resulting from the large error bars.  The ultimate affect is the probability distribution is stretched.  We therefore consider models summing to 68\%~of the total probability (e.g., 19,000--27,000~K) to be the maximum range of models that can fit the COS spectrum.

\subsection{Other Sources of FUV Flux}
\label{sec:other}

While we do find, like \citet{maund04}, that the surrounding stars contribute little to the overall flux, we also consider the possibility that this hot B-star emission may not arise from the supposed binary companion.  First, we cannot definitively rule out line-of-sight coincidences.  Moreover, the high-resolution TA image contours in Figure \ref{fig4} reveal the presence of an unresolved UV source just south of the SN position.  We also see a hint of this feature in the WFC/F336W filter in Figure \ref{fig1}.  At a distance of 3.6 Mpc, this angular separation of $> 0.1$\arcsec\ corresponds to $> 1.5$~pc, thereby ruling out the possibility of this source being the putative companion of the progenitor.  Furthermore, the flux of this unresolved source is low relative to the SN+putative companion.  The presence of this emission, however, suggests the possibility that the local progenitor environment is crowded.  If this is true, the B2-star detection within the $HST$~PSF does not necessarily have to be the binary companion.  Future {\it HST} imaging may be able to probe the progenitor environment in more detail once the SN sufficiently fades.

%%Error Distribution

\section{Conclusion and Future Work}
\label{sec:con}

We present COS spectra and WFC3 photometry of SN 1993J nearly 20~yr post-explosion.  The SN emission has faded sufficiently that we can test predictions of the presence of the progenitor's companion.  The rise in the NUV, previously explained by the presence of the hot star, is still consistent with the scaled spectrum of SN 1993J obtained by {\it HST}/STIS in the year 2000.  The continued rise in the FUV, however, cannot be explained by this model, even with flux contributions from neighboring stars that contaminate the spectrum.  The addition of a hot B-star (B3--B1) spectrum most accurately accounts for this FUV excess.  

The properties of this companion can be used to directly address several questions.  The temperature, surface gravity, and bolometric luminosity can be used to constrain the original stellar mass \citep[e.g.,][]{claeys11}.  The B2 star binary companion scenario is also interesting when considering the SN 1993J circumstellar medium (CSM), which has a flattened or disk-like geometry \citep{matheson00}, and derived progenitor mass-loss rates of $10^{-4}$--$10^{-5}$~\ml\ for a wind velocity of 10~\kms~\citep[e.g.,][]{dyk94,fransson96,mioduszewski01,immler01,fransson05,chandra09}.  B2~Ia stars typically have mass-loss rates of (0.5--1) $\times 10^{-6}$~\ml\ and a wind velocity of $\sim 500$~\kms\ \citep{crowther06}.  The fact that the CSM density of SN 1993J is three orders of magnitudes higher than that of typical B2 stars implies that the SN 1993J CSM was formed by the primary star that exploded, or during Roche-lobe overflow (RLOF) prior to the explosion.  Such a system is relevant in the context of several nearby B supergiants surrounded by ring-like nebulae, including the B3~I progenitor to SN~1987A \citep[e.g.,][]{burrows95, smith07b}, Sher~25 (a B1.5~Ia star; \citealt{brandner97}; \citealt{smith07b}), HD 168625 \citep{smith07a}, and SBW1 (a B1.5~Ia star; \citealt{smith07b,smith13}).  Many of these systems have previously been considered products of RLOF from a companion, products of mergers, mass loss from a rapidly rotating star, or a stripped-envelope mass loser that has already exploded as a SN~Ibc or IIb, such as SN 1993J.   None of these possibilities has yet been confirmed.

While this paper disentangles the NUV contributions of nearby stars and the underlying SN, the FUV remains relatively unconstrained.  Now that the SN flux has faded sufficiently, future FUV/NUV imaging of SN 1993J is necessary to confirm the presence of a hot B star and constrain the FUV contribution from these sources.\\

\vspace{-0.1in}
\acknowledgments

This work is based on observations made with the NASA/ESA {\it Hubble Space Telescope}, obtained from the Space Telescope Science Institute (STScI), which is operated by the Association of Universities for Research in Astronomy (AURA), Inc., under NASA contract NAS5-26555. Financial support was provided by NASA through grant GO-12531 from STScI, NSF grant AST-1211916, the TABASGO Foundation, and the Christopher R. Redlich Fund. The research by C.F. is supported by the Swedish Research Council and National Space Board. We are grateful to the STScI Help Desk for their assistance with the $HST$~data. Some of the data presented herein were obtained at the W.~M. Keck Observatory, which is operated as a scientific partnership among the California Institute of Technology, the University of California, and NASA; the observatory was made possible by the generous financial support of the W.~M. Keck Foundation. We thank the staff of the Keck Observatory, together with Kelsey Clubb, WeiKang Zheng, and Adam Miller, for their assistance with the observations.  Selma de Mink provided enlightening discussions.  

{\it Facilities:} \facility{HST(WFC3)}, \facility{HST(COS)}, \facility{Keck (LRIS)}

\bibliographystyle{apj}
\bibliography{references}

\end{document}